\documentclass[journal,twoside,final]{IEEEtran}

\usepackage{graphicx}
\usepackage{subfigure}
\usepackage{amsmath}
\usepackage{amsfonts}
\usepackage{amssymb}
\usepackage{stfloats}
\usepackage[latin1]{inputenc}
\usepackage[T1]{fontenc}
\usepackage[no adjust]{cite}
\usepackage{url}
\usepackage{comment}

\usepackage{amsfonts}
\usepackage{bbm}
\usepackage{mathrsfs}
\usepackage{graphicx}
\usepackage{amsmath}
\usepackage{amssymb}
\usepackage{amsbsy}
\usepackage{amstext}
\usepackage{amsopn}
\usepackage{amsthm}
\usepackage{amssymb}
\usepackage{eufrak}
\usepackage{cite}
\usepackage{fancybox}
\usepackage{dashbox}
\usepackage{slashbox}
\usepackage{threeparttable}

\newtheorem{theorem}{Theorem}
\newtheorem{definition}{Definition}
\newtheorem{lemma}{Lemma}

\newtheorem{remark}{Remark}
\def\QEDclosed{\mbox{\rule[0pt]{1.3ex}{1.3ex}}} %

\def\QED{\QEDclosed}

\newcommand{\tqbinom}[2]{\genfrac{[}{]}{0pt}{}{#1}{#2}}
\newenvironment{ventry}[1]%
{\begin{list}{}
{ %
\settowidth{\labelwidth}{ \textsf{#1} } %
\setlength{\labelsep}{3mm}
\setlength{\leftmargin}{\labelwidth+\labelsep}}} %
{\end{list}}

\usepackage{flushend}
\begin{document}
\title{\LARGE Block-Orthogonal Space-Time Code Structure and Its Impact on QRDM Decoding Complexity Reduction}
\author{Tian Peng Ren, Yong Liang Guan, Chau Yuen, Er Yang Zhang

\thanks{
Manuscript received Jan 25, 2011; revised Jun 15, 2011. The work of C. Yuen was supported by the International Design Center (Grant No. IDG31100102 \& IDD11100101). The material in this paper was partly presented at the IEEE SPAWC, Marrakech, Morocco, June 20-23, 2010 and IEEE VTC'10-Fall, Ottawa, Canada, 6-9 Sept. 2010.}

\thanks{
T. P. Ren and E. Y. Zhang are with the College of Electronic Science and Engineering, National University of Defense Technology, Changsha 410073, China (e-mail: tpren@nudt.edu.cn; eyzhang2006@hotmail.com). T. P. Ren is now with 63790 troops, Xichang 615000, China.}

\thanks{
Y. L. Guan is with the School of Electrical and Electronic Engineering, Nanyang Technological University, Singapore 639798 (e-mail: eylguan@ntu.edu.sg).}

\thanks{
C. Yuen is with Singapore University of Technology and Design, Singapore 279623(e-mail: yuenchau@sutd.edu.sg).}

\thanks{
Color versions of one or more of the figures in this paper are available online at http://ieeexplore.ieee.org.}

\thanks{
Digital Object Identifier 10.1109/JSTSP.2011.2166755}

}
\maketitle

\markboth{IEEE JOURNAL OF SELECTED TOPICS IN SIGNAL PROCESSING, VOL. 5, NO. 8, DECEMBER 2011}{REN \MakeLowercase{\textit{et al.}}: BLOCK-ORTHOGONAL SPACE-TIME CODE STRUCTURE AND ITS IMPACT ON QRDM DECODING COMPLEXITY REDUCTION}

\pubid{1932-4553/\$26.00~\copyright~2011 IEEE}

\pubidadjcol

\begin{abstract}
Full-rate space time codes (STC) with rate = \emph{number of transmit antennas} have high multiplexing gain, but high decoding complexity even when decoded using reduced-complexity decoders such as sphere or QRDM decoders. In this paper, we introduce a new code property of STC called \emph{block-orthogonal} property, which can be exploited by QR-decomposition-based decoders to achieve significant decoding complexity reduction without performance loss. We show that such
complexity reduction principle can benefit the existing algebraic codes such as Perfect and DjABBA codes due to their inherent (but previously undiscovered) block-orthogonal property. In addition, we construct and optimize new full-rate BOSTC (Block-Orthogonal STC) that further maximize the QRDM complexity reduction potential. Simulation results of bit error rate (BER) performance against decoding complexity show that the new BOSTC outperforms all previously known codes as long as the QRDM decoder operates in reduced-complexity mode, and the code exhibits a desirable complexity saturation property.
\end{abstract}

\IEEEpeerreviewmaketitle

\section{Introduction}
\PARstart{B}ecause of their simple maximum-likelihood (ML) decoding, space-time codes (STC) with pure orthogonal property have received considerable attention in the past decade \cite{Alamouti,Tarokh_ostbc,Ganesan_ostbc,Lu_ostbc}. However, the code rates of orthogonal STC are mostly low\cite{Wang_upperbound}. To increase the code rates, pure orthogonality has been relaxed to quasi-orthogonality for STC in \cite{qostbc_Jafarkhani,qostbc_Dao,qostbc_Mecklenbrauker,qostbc_Papadias,qostbc_Rajan,qostbc_Tirkkonen,
qostbc_Yuen,2gp_Yuen,2gp_Rajan,2gp_Ren}.

To pursue high transmission rates, high-rate STC such as Bell Labs layered space-time (BLAST) \cite{Foschini}, double space-time transmit diversity (D-STTD) code \cite{Texas_dsttd}, DjABBA code \cite{Hottinen} and algebraic STC \cite{Belfiore}\cite{Oggier} have been developed, but they demand a high maximum-likelihood (ML) decoding complexity\footnote{In this paper, decoding complexity represents the number of likelihood function calculations per symbol duration in a decoding process.} due to the non-orthogonal code structure. In order to reduce the decoding complexity of existing algebraic STC, fast-decodable structure is proposed in \cite{Biglieri}, however, the associated complexity reduction is upper bounded by the maximum code rate of (quasi-)orthogonal STC \cite{Ren_fgd}, and hence is limited.

Basically, quasi-orthogonality and fast-decodability in STC imply additional zero entries in the upper triangle matrix after QR decomposition of the equivalent channel matrices, these zero entries are exploited in breadth-first search or depth-first search decoders such as QRDM and sphere decoders to achieve decoding complexity reduction. In this paper, we introduce a new property for full rate STC, called \emph{block-orthogonal} property, and propose further QRDM complexity reduction for codes with such property. The proposed decoding principle can benefit many existing algebraic codes due to their previously undiscovered block-orthogonal property. For example, D-STTD code and DjABBA code have about 50\% decoding complexity reduction. Moreover, we design new full-rate codes called block orthogonal STC (BOSTC) that further exploit the block-orthogonal property for complexity reduction in QRDM decoders. Besides the usual bit error rate (BER) against signal-to-noise ratio (SNR)investigation approach, we also adopt a new approach: BER comparison against decoding complexity, which gives interesting new insights into codes which are optimal with respect to specific decoding complexity levels.

The rest of this paper is organized as follows. System model is presented in Section \ref{sec_systemmodel}. Block-orthogonal property and BOSTC are introduced and studied in Section \ref{sec_BOSTC}. In Section \ref{sec_bostc_benefit}, the benefit of block-orthogonal property is described and simulated. New BOSTC for arbitrary transmit antenna number are constructed and optimized in Section \ref{sec_bostcconstruct}. The bit error rate (BER) performance simulations are provided in Section \ref{sec_Simulation}. This paper is concluded in Section \ref{sec_conclusion}.

In what follows, bold lower case and upper case letters denote vectors and matrices (sets), respectively; $\mathbb{R}$ and $\mathbb{C}$ denote the real and the complex number field, respectively; $(\cdot)^R$ and $(\cdot)^I$ stand for the real and the imaginary part of a complex vector or matrix, respectively; $[\cdot]^T$, $[\cdot]^H$, $| \cdot |$ and $rank(\cdot)$ denote the transpose, the complex conjugate transpose, the Frobenius norm and the rank of a matrix, respectively; $[a_{ij}]$ denotes a matrix with the $i$-th row and the $j$-th column element $a_{ij}$.

\section{System Model}\label{sec_systemmodel}
\subsection{Signal Model}
We consider a space-time coded multi-input multi-output (MIMO) system employing $N_t$ transmit antennas and $N_r$ receive antennas. Let the transmitted signal sequences be partitioned into independent time block, denoting as $\{s_1,s_2,$

\newpage\noindent$\cdots,s_L\}$ where $s_l$ are real-valued information symbols\footnote{The in-phase component or the quadrature component of a complex information symbol is real, hence, this signal model is also applicable for complex information symbol transmission.} for transmission. To transmit $\{s_1,s_2,\cdots,s_L\}$ from $N_t$ transmit antennas over $T$ symbol durations, an STBC matrix $\textbf{X} \in \mathbb{C}^{T\times N_t}$ is designed following the signal model in \cite{Hassibi}:
\begin{equation}\label{eq_LSTBC}
\textbf{X}=\sum^L_{l=1}{s_{l}\textbf{C}_{l}}
\end{equation}
\noindent where $\textbf{C}_l\in \mathbb{C}^{T\times N_t}~(l=1,\cdots,L)$ are called dispersion matrices. The code rate is $\frac{L}{2T}$ considering complex symbol transmission, and the average energy of the code matrix $\textbf{X}$ is constrained to $\mathcal {E}_\textbf{X}=\mathbb{E}\|\textbf{X}\|^2=T$.

The received signals $\tilde{y}_{tm}$ of the $m$th $(m=1,\cdots,N_r)$ receive antenna at time $t$ $(t=1,\cdots,T)$
can be arranged in a $T\times {N_r}$ matrix $\tilde{\textbf{Y}}=\left[\tilde{\textbf{y}}_1~\tilde{\textbf{y}}_2~\cdots~\tilde{\textbf{y}}_{N_r}\right]$.
Thus, the transmit-receive signal relation can be represented as:
\begin{equation}
\tilde{\textbf{Y}}=\sqrt{\rho}\textbf{X}\tilde{\textbf{H}}+\tilde{\textbf{Z}}
\end{equation}
\noindent where $\tilde{\textbf{H}}_{N_t\times {N_r}} =
\left[\tilde{\textbf{h}}_1~\tilde{\textbf{h}}_2~\cdots~\tilde{\textbf{h}}_{N_r}\right]$
is the channel coefficient matrix. We often assume that the communication channel is quasi-static Rayleigh fading with coefficient of independently, identically distributed (i.i.d.) $\mathcal {CN}(0,1)$ entries; $\tilde{\textbf{Z}}_{T\times {N_r}}
=\left[\tilde{\textbf{z}}_1~\tilde{\textbf{z}}_2~\cdots~\tilde{\textbf{z}}_{N_r}\right]=
\left[\tilde{z}_{tm}\right]$ is the additive white Gaussian noise (AWGN) matrix where the entries $\tilde{z}_{tm}$ are independently, identically distributed (i.i.d.) $\mathcal {CN}(0,1)$; $\rho$ is the average SNR at each receive antenna.

Following the signal model in \cite{Hassibi}, the received signal can also be shown to be:
\begin{equation}\label{eq_r_Hs}
\textbf{y}=\sqrt{\rho}\textbf{H}\textbf{s}+\textbf{z}
\end{equation}
with $l=1,2,\cdots,L$ and

\begin{equation*}
\begin{split}
 &\textbf{y}=\left[
\begin{array}{cccccccc}
    \tilde{\textbf{y}}^R_1\\
    \tilde{\textbf{y}}^I_1\\
    \vdots\\
    \tilde{\textbf{y}}^R_{N_r}\\
    \tilde{\textbf{y}}^I_{N_r}
\end{array}
\right],\bar{\textbf{h}}=\left[
\begin{array}{cccccccc}
    \tilde{\textbf{h}}^R_{1}\\
    \tilde{\textbf{h}}^I_{1}\\
    \vdots \\
    \tilde{\textbf{h}}^R_{N_r}\\
    \tilde{\textbf{h}}^I_{N_r}
\end{array}
\right],\textbf{s}=\left[
\begin{array}{cccccccc}
    s_1\\
    s_1\\
    \vdots\\
    s_L
\end{array}
\right],\textbf{z}=\left[
\begin{array}{cccccccc}
    \tilde{\textbf{z}} ^R_1\\
    \tilde{\textbf{z}} ^I_1\\
    \vdots\\
    \tilde{\textbf{z}} ^R_{N_r}\\
    \tilde{\textbf{z}} ^I_{N_r}
\end{array}
\right]
\end{split}
\end{equation*}

\begin{equation*}
\begin{split}
&
\textbf{H}=[\textbf{h}_1,\textbf{h}_2,\cdots,\textbf{h}_{L}]=\left[
\begin{array}{cccccccc}
    \mathscr{C}_1\bar{\textbf{h}} &\mathscr{C}_2\bar{\textbf{h}}  &\cdots  &\mathscr{C}_L\bar{\textbf{h}}
\end{array}
\right]
\end{split}
\end{equation*}
\begin{equation*}
\begin{split}
\mathscr{C}_l=\left[
\begin{array}{cccccccc}
    \mathcal {C}_l   &   \textbf{0} & \cdots & \textbf{0}\\
    \textbf{0}   &   \mathcal {C}_l & \cdots & \textbf{0}\\
    \vdots & \vdots & \ddots & \vdots \\
    \textbf{0} & \textbf{0} & \cdots & \mathcal {C}_l
\end{array}
\right]_{N_r\times N_r},\mathcal {C}_l=\left[
\begin{array}{cccccccc}
    \textbf{C}^R_l   &   -\textbf{C}^I_l\\
    \textbf{C}^I_l   &   \textbf{C}^R_l
\end{array}
\right]_{2\times 2}
\end{split}
\end{equation*}
\noindent where $\textbf{y}\in \mathbb{R}^{2TN_r\times 1},~\textbf{s}\in
\mathbb{R}^{{L}\times 1},~\textbf{z}\in \mathbb{R}^{2TN_r\times 1}$
and $\textbf{H}\in \mathbb{R}^{2TN_r\times {L}}$ are the equivalent received signal vector, information symbol vector, equivalent noise vector and equivalent channel matrix, respectively.

To avoid rank deficiency at the decoder, $rank(\textbf{H})=L$ is
required, which means that $\textbf{H}$ should be \textquoteleft
\textquoteleft tall\textquoteright\textquoteright, i.e., $L\leq
2TN_r$ \cite{Hassibi}\cite{Yuen_book}. Therefore, we assume that the
number of receiver antennas $N_r\geq \frac{L}{2T}$. Moreover, $\tqbinom{\textbf{C}^R_1}{\textbf{C}^I_1}$,
$\tqbinom{\textbf{C}^R_2}{\textbf{C}^I_2},\cdots$, $\tqbinom{\textbf{C}^R_{L}}{\textbf{C}^I_{L}}$ must be linearly
independent to guarantee $rank(\textbf{H})=L$ \cite{2gp_Ren}.

\subsection{Code Rate of STC}

\begin{lemma}\cite{Hassibi}\cite{Yuen_book} \label{prop_maxrate}
\emph{In an $N_t\times N_r$ MIMO system, the code rate of STC applied cannot exceed the
minimum of transmit and receive antenna numbers, i.e.,
\begin{equation}
Rate=\frac{L}{2T}\leq \min(N_t,N_r).
\end{equation}}
\end{lemma}

\begin{definition}[Full-Rate STC]\label{definition_fullrate}
\emph{An STC for $N_t\times N_r$ MIMO systems is full-rate when its code rate achieves the value of $\min(N_t,N_r)$.
~~~~~~~~~~~~~~~~~~~~~~~~~~~~~~~~~~~~~~~~~~~~~~~~~~~~~~~~~~\QED
}
\end{definition}

In this paper, we always assume that $N_t\leq N_r$, hence an STC is full-rate when its code rate achieves the value of $N_t$.

\section{Block-Orthogonal STC}\label{sec_BOSTC}
In this section, block-orthogonal STC (BOSTC) and block-orthogonal code property \cite{ren_bostc} are defined and discussed.

\subsection{Definition of BOSTC }
Most reduced-complexity MIMO decoders such as sphere decoder \cite{Damen} and QR decoder with M-algorithm (QRDM)
\cite{Kim}\cite{Chin} are based on QR decomposition. With QR decomposition, BOSTC is defined as follows:
\begin{definition}[BOSTC]\label{def_bostbc1}
\emph{Suppose that $\textbf{H}_{2TN_r\times L}$ is the equivalent channel matrix when an STC $\textbf{X}_{T\times N_t}$ is applied in $N_t\times N_r$ MIMO systems. Denoting QR decomposition on $\textbf{H}$ as: $\textbf{H}=\textbf{QR}$ where
$\textbf{Q}=\left[\textbf{q}_1~\cdots~\textbf{q}_{L}\right]\in \mathbb{R}^{2TN_r\times L}$ is unitary and $\textbf{R}\in \mathbb{R}^{L\times L}$ is upper-triangular, $\textbf{X}$ is called block-orthogonal STC (BOSTC) and have block-orthogonal structure if}
\emph{
\begin{equation} \label{eq_R_bostc}
\begin{split}
\textbf{R}=\left[
\begin{array}{cccccccc}
    \textbf{D}_1   &   \textbf{E}_{12}         & \cdots   &   \textbf{E}_{1\Gamma}\\
    \textbf{0}     &   \textbf{D}_2   & \cdots   &   \textbf{E}_{2\Gamma}\\
    \vdots        &   \vdots         & \ddots   &   \vdots \\
    \textbf{0}    &   \textbf{0}     & \cdots   &   \textbf{D}_{\Gamma}\\
\end{array}
\right]
\end{split}
\end{equation}
}
\emph{\noindent where the sub-block $\textbf{D}_i$ is full-rank diagonal matrix of size $k_i\times k_i$ as shown in (\ref{d_bostc}), and the information symbols corresponding to the same sub-block are independent (i.e., their values represent independent information) and orthogonal (i.e., their dispersion matrices satisfy the quasi-orthogonal constraints (QOC) in \cite{qostbc_Yuen}); $\Gamma$ is the number of sub-blocks $\textbf{D}$'s and $\sum_{i=1}^{\Gamma}{k_i}=L$; $\textbf{E}_{i_1i_2}(i_1=1,2,\cdots,\Gamma-1,~i_2=i_1+1,\cdots,\Gamma)$ denotes matrix containing arbitrary values.~~~~~~~~~~~~~~~~~~~~~~~~~~~~~~~~~~~~~~~~~~~~~~~~~~~~~~~~~~~~~~~~~~~~~~~~~~~\QED }
\emph{\begin{equation}\label{d_bostc}
\begin{split}
\textbf{D}_i=
diag \left(u_{i,1},~u_{i,2},~\cdots,~u_{i,k_i}\right)
\end{split}
\end{equation}}
\end{definition}

In Def. \ref{def_bostbc1}, $\textbf{D}_i~(i=1,2,\cdots,\Gamma)$ in (\ref{d_bostc}) are diagonal matrices with non-zero scalar diagonal entries. If these scalar diagonal entries are replaced with square upper-triangular matrices such as:
\begin{equation}\label{eq_d_bqosmc}
\begin{split}
\textbf{D}_i=
diag \left(\textbf{U}_{i,1},~\textbf{U}_{i,2},~\cdots,~\textbf{U}_{i,{k_i}} \right)
\end{split}
\end{equation}
\noindent where $\textbf{U}_{i,k}$ are full-rank upper-triangular matrices of
size $\gamma_{i,{k}}\times \gamma_{i,{k}}$ with
$\sum^\Gamma_{i=1}\sum^{k_i}_{\kappa=1}{\gamma_{i,{\kappa}}}=L$,
$i=1,2,\cdots,\Gamma$, $\kappa=1,2,\cdots,{k_i}$, then the code can
be viewed as a \textit{block-quasi-orthogonal code} (instead of
block orthogonal). The information symbols corresponding to the same
sub-block $\textbf{D}$ and different $\textbf{U}$s are independent
and orthogonal.

In general, the size of (block-)diagonal matrices $\textbf{D}$'s and upper-triangular matrices $\textbf{U}$'s can be arbitrary. In this paper only the case that $\textbf{D}$'s have the same size $k\times k$ (i.e., $k_1=k_2=\cdots =k_{\Gamma}\triangleq k$) and $\textbf{U}$'s have the same size $\gamma\times \gamma$ (i.e., $\gamma_{1,{1}}=\cdots=\gamma_{1,{k}}=\cdots =\gamma_{\Gamma,1}=\cdots =\gamma_{\Gamma,{k}}\triangleq \gamma$) is considered. Hence, block-(quasi-)orthogonal structure can be unified by three parameters as $(\Gamma, k, \gamma)$:
\begin{ventry}{$\bullet$}
\item[$\bullet$]$\Gamma$: the number of matrices $\textbf{D}$ (i.e., sub-blocks) in $\textbf{R}$;
\item[$\bullet$]$k$: the number of scalars $u$ or matrices $\textbf{U}$ in $\textbf{D}$'s;
\item[$\bullet$]$\gamma$: the number of diagonal entries in matrices $\textbf{U}$ ($\gamma=1$ for scalars $u$ ).
\end{ventry}

To simplify the notations further, in the sequel of this paper we will not make distinction between block-orthogonal STC and block-quasi-orthogonal STC. They will both be called \textquoteleft
\textquoteleft block-orthogonal STC\textquoteright\textquoteright with parameters $(\Gamma, k, \gamma)$.

\subsection{Block-Orthogonal Property}
In this section, we present sufficient conditions for an STC to attain block-orthogonal structure.
\subsubsection{2-Block BOSTC}
We first propose a sufficient condition for an STC to achieve block-orthogonal structure $(\Gamma=2,k,\gamma=1)$. The case of $\Gamma>2$ will be discussed subsequently.
\begin{theorem}\label{th_bostbc_a}
\emph{Considering an STC of size $T\times N_t$ with dispersion matrices $\textbf{A}_1,\cdots, \textbf{A}_k,~\textbf{B}_{1},\cdots, \textbf{B}_{k}$\footnote{For ease of presentation, here we employ $\{\textbf{A}\}$ and $\{\textbf{B}\}$ as dispersion matrices, instead of $\{\textbf{C}\}$ presented in (\ref{eq_LSTBC}).}. Let
\begin{equation*}
\begin{split}
\mathcal {A}_i=\left[
\begin{array}{cccccccc}
    \textbf{A}^R_i   &   -\textbf{A}^I_i \\
    \textbf{A}^I_i   &    \textbf{A}^R_i
\end{array}
\right],~ \mathcal {B}_i=\left[
\begin{array}{cccccccc}
    \textbf{B}^R_i   &   -\textbf{B}^I_i \\
    \textbf{B}^I_i   &    \textbf{B}^R_i
\end{array}
\right]
\end{split}
\end{equation*}
and $\mathcal {A}_i\triangleq \left[ a_{iup}\right]_{2T\times
2N_t}$, $\mathcal {B}_i\triangleq \left[ b_{iup}\right]_{2T\times
2N_t}(i=1,\cdots,k,~u=1,\cdots,2T,~p=1,\cdots,2N_t)$, then this STC
has block-orthogonal structure $(2,k,1)$ if
\begin{subequations}\label{bostc_condition}
\begin{equation}\label{bostc_condition1}
\begin{split}
1.~\{\mathcal{A}_1,\cdots,\mathcal{A}_k,\mathcal{B}_1,\cdots,\mathcal{B}_k\}\emph{\emph{
is of dimention
}}2k;~~~~
\end{split}
\end{equation}
\begin{equation}\label{bostc_condition2}
\begin{split}
2.~\mathcal{A}^T_i\mathcal{A}_i=\textbf{I},~\mathcal{B}^T_i\mathcal{B}_i=\textbf{I}~
(i=1,\cdots,k);~~~~~~~~~~~~~~
\end{split}
\end{equation}
\begin{equation}\label{bostc_condition3}
\begin{split}
3.~\mathcal{A}^T_i\mathcal{A}_j=-\mathcal{A}^T_j\mathcal{A}_i~(i,j=1,\cdots,k\emph{\emph{
and }}i\neq
j);~~~~~
\end{split}
\end{equation}
\begin{equation}\label{bostc_condition4}
\begin{split}
4.~\mathcal{B}^T_i\mathcal{B}_j=-\mathcal{B}^T_j\mathcal{B}_i~(i,j=1,\cdots,k\emph{\emph{
and }}i\neq
j);~~~~~~
\end{split}
\end{equation}
\begin{equation}\label{bostc_condition5}
\begin{split}
& 5.~\sum_{(p,q,s,t)\in \mathbb{S}} d_{pqst}=0 (~i,j=1,\cdots,k
,\text{and}~i\neq j) \\& \emph{\emph{ where }}
d_{pqst}=\sum^{k}_{\kappa=1}\left( \sum^{2T}_{u=1}b_{iup}a_{\kappa
us}\cdot\sum^{2T}_{v=1}b_{jvq}a_{\kappa vt}\right).~~~
\end{split}
\end{equation}
\end{subequations}
each element (tuple) of set $\mathbb{S}$ includes 4
uniquely-permuted scalars\footnote{For example, $\sum_{(1,2,1,1)\in
\mathbb{S}} d_{pqst}= d_{1112}+d_{1121}+d_{1211}+d_{2111}$ and
$\sum_{(1,2,3,1)\in \mathbb{S}} d_{pqst}=
d_{1123}+d_{1132}+d_{1213}+d_{1312}+d_{1231}+d_{1321}+d_{2113}+d_{2131}+d_{2311}+d_{3112}+d_{3121}+d_{3211}$.}
drawn from $\{1,\cdots,2N_t\}$.~~~~~~~~~~~~~~~~~~~~~~~~~~~~~~~~~~~~~~~~\QED }
\end{theorem}

The proof of Theorem \ref{th_bostbc_a} is given in Appendix \ref{proof_th_bostbc_a}. Based on Theorem \ref{th_bostbc_a}, the $2\times2$ fast-decodable codes in \cite{Paredes,Sezginer,Rabiei} can be shown to have block-orthogonal structure $(2,4,1)$.

\vspace{0.08in}
\subsubsection{$\Gamma$-Block BOSTC ($\Gamma>2$)}
\begin{definition} \label{def_borelation}
\emph{Consider an STC with dispersion matrices
$\{\textbf{A}_1,\cdots, \textbf{A}_\mathbbm{k}\}$ and
$\{\textbf{B}_1,\cdots,\textbf{B}_k\}$ and an associated equivalent
channel matrix $\textbf{H}$, the matrices $\{\textbf{B}_{1},\cdots,
\textbf{B}_{k}\}$ is said to satisfy block QOC (i.e., conditions (8b) and (8c)) under matrices
$\{\textbf{A}_1,\cdots, \textbf{A}_\mathbbm{k}\}$ if their
associated
$\textbf{R}(\mathbbm{k}+1:\mathbbm{k}+k,\mathbbm{k}+1:\mathbbm{k}+k)$
is diagonal, where $\textbf{R}$ is the upper-triangular matrix after
QR decomposition of $\textbf{H}$,
$\textbf{R}(\mathbbm{k}+1:\mathbbm{k}+k,\mathbbm{k}+1:\mathbbm{k}+k)$
is the sub-matrix constituted by the $(\mathbbm{k}+1)$th to
$(\mathbbm{k}+k)$th rows and the $(\mathbbm{k}+1)$th to
$(\mathbbm{k}+k)$th columns of
$\textbf{R}$.~~~~~~~~~~~~~~~~~~~~~~~~~~~~~~~~~~~~~~~~~~~~~~~~~~~~~~~~~~~~~~~~~~~~~~~~~~~~~\QED
}
\end{definition}

Based on Def. \ref{def_borelation}, a sufficient condition to check whether an STC has block-orthogonal structure $(\Gamma,k,1)$ is provided as follows.
\begin{theorem}\label{th_bostbc_b}
\emph{Denoting the equivalent channel matrix of an STC with
dispersion matrices $\{\textbf{A}_1,\cdots,
\textbf{A}_\mathbbm{k}\}$ and $\{\textbf{B}_1,\cdots,\textbf{B}_k\}$
as $\textbf{H}=\left[\textbf{H}_1~\textbf{H}_2\right]$,
$\textbf{H}_1=\left[\textbf{h}_1~\cdots~\textbf{h}_{\mathbbm{k}}\right]$,
$\textbf{H}_2=\left[\textbf{h}_{\mathbbm{k}+1}~\cdots~\textbf{h}_{\mathbbm{k}+k}\right]$,
the matrices $\{\textbf{B}_{1},\cdots, \textbf{B}_{k}\}$ satisfy
block-QOC under matrices $\{\textbf{A}_1,\cdots,
\textbf{A}_\mathbbm{k}\}$ if}

\emph{1) $\{\textbf{B}_{1},\cdots, \textbf{B}_{k}\}$ satisfy the
QOC;}

\emph{2) the projection coefficient matrix of vectors
$\textbf{h}_{\mathbbm{k}+1},\cdots,\textbf{h}_{\mathbbm{k}+k}$ onto
vector space $\{\textbf{h}_1,\cdots,\textbf{h}_{\mathbbm{k}}\}$,
i.e., the sub-matrix formed by the first column to the
$\mathbbm{k}$-th column and the $(\mathbbm{k}+1)$-th row to the
$(\mathbbm{k}+k)$-th row of $\textbf{B}$ in (\ref{eq_R_bostc}), is
para-unitary\footnote{$\textbf{A}$ is para-unitary if
$\textbf{A}^H\textbf{A}=\textbf{I}$.}.~~~~~~~~~~~~~~~~~~~~~~~~~~~~~~~~~~~~~~~~~~~~~~~~~~~~~~~~~~\QED}
\end{theorem}
The proof of Theorem \ref{th_bostbc_b} is given in Appendix \ref{proof_th_bostbc_b}. Using Theorems \ref{th_bostbc_b}, we can classify the block-orthogonal structure achieved by many existing codes, as shown in Table \ref{table_existingcode}. We can see that all these codes have block-orthogonal structure with $k\leq2$.
\begin{table*}[!t]
\begin{center}
\begin{threeparttable}[!t]
\tabcolsep 7mm
\caption{Block-Orthogonal Structure of Existing Codes for $N_t$ Transmit Antennas over $T$ Symbol
Durations$^a$.} \label{table_existingcode}
\newcommand{\rb}[1]{\raisebox{2.2ex}[0pt]{#1}}
\newcommand{\rbb}[1]{\raisebox{1.0ex}[0pt]{#1}}
{\small\begin{tabular}{|c|c|c|c|c|c|} \hline
{\rule[-0mm]{0mm}{4mm}}  &  $N_t$ & $T$ & $\Gamma$ & $k$ & $\gamma$
\\\hline\hline
{\rule[-0mm]{0mm}{4mm}}BLAST\cite{Foschini}        &$N_t$ & 1 &$N_t$
& {2} & 1
\\ \hline
{\rule[-0mm]{0mm}{4mm}}Golden code\cite{Belfiore}  & 2 & 2   & 4 &
{2} & 1
\\ \hline
{\rule[-0mm]{0mm}{4mm}}D-STTD code\cite{Texas_dsttd} & 4 & 2 & 2 & 4 & 1
\\ \hline
{\rule[-0mm]{0mm}{4mm}}DjABBA code\cite{Hottinen}  & 4 & 4   & 4 &
{2} & 2
\\ \hline
{\rule[-0mm]{0mm}{4mm}} & 3 (or 6) &3 (or 6)   & 9 (or 36) & {1} & 2
\\ \cline{2-6}
{\rule[-0mm]{0mm}{4mm}}\rb{Perfect code\cite{Oggier}} & 4 & 4 & 16 &
{2} & 1
\\ \hline
\end{tabular}
\begin{tablenotes}
\item[$^a$]Without special requirements (e.g., to achieve full diversity, constellation rotation is required for DjABBA code \cite{Hottinen} and HEX constellation is applied for Perfect code with 3 (or 6) transmit antennas \cite{Oggier}), we assume that each complex information symbol is drawn from a square QAM without constellation rotation, equivalently, each real information symbol is drawn from a one-dimension constellation.
\end{tablenotes}
}
\end{threeparttable}
\end{center}
\hrulefill \vspace*{0pt}
\end{table*}

\subsection{Relationship with the Existing Works}
\begin{figure*}[!t]
\centering
\includegraphics[width=5.2in]{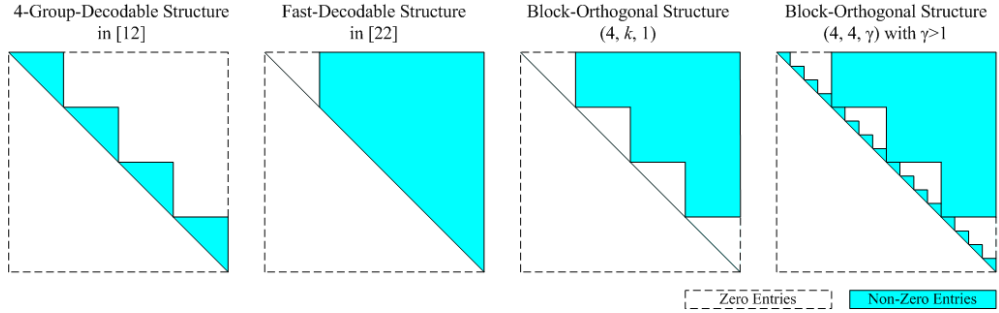}
\caption{$\textbf{R}$'s of group-decodable structure, fast-decodable
structure and block-orthogonal structure.} \label{gp_fd_bo_fig}
\end{figure*}

The $\textbf{R}$'s after the QR decomposition on the equivalent
channel matrices $\textbf{H}$'s of group-decodable, fast-decodable
structure and block-orthogonal STC are compared graphically in Fig.
\ref{gp_fd_bo_fig}.
The group decodable codes can not achieve full code rates
\cite{2gp_Ren}. On the other hand, the fast-decodable
codes\cite{Biglieri} have very few zeros in the $\textbf{R}$ matrix,
hence limited decoding complexity reduction. This is the reason why
we introduce full-rate block-orthogonal structure in this paper.

\section{Benefit of Block-Orthogonal Structure: Decoding Complexity Reduction}\label{sec_bostc_benefit}
In this section, we first review the breadth-first search decoding used in \emph{traditional} QRDM \cite{Kim}\cite{Chin}, and then propose a \emph{simplified} QRDM that exploits the block-orthogonal structure. Next, we use the D-STTD code with block-orthogonal structure $(2,4,1)$ to illustrate the decoding complexity reduction.

Assume that the real information symbols corresponding to an upper-triangular matrix $\textbf{U}$ are drawn from a constellation of size $M$. For example, if each real information symbol in an STC with block-orthogonal structure $(\Gamma,k,\gamma)$ is 4-PAM (pulse amplitude modulation) modulated, the symbols corresponding to the same $\textbf{U}$ are drawn from a constellation of size $M=4^\gamma$.

\subsection{Traditional QRDM}
In traditional QRDM, the surviving paths with smaller accumulated Euclidean distance (Euclidean metric) are picked from the full (ML decoding) or partial(near-ML decoding) search tree. Assume that at each stage $M_c$ paths are reserved, as shown in Fig. \ref{fig_qrdv} where $M_c=3$. At the beginning, all the search paths are reserved until the number of total search paths exceeds $M_c$; then only $M_c$ paths with the smallest accumulated Euclidean metrics, surviving paths, are picked for the Euclidean metric calculations in the next stage. In this case, $MM_c$ metrics need to be calculated in each stage. Hence, the decoding complexity (likelihood function calculation number) under traditional QRDM is:
\begin{equation}\label{eq_normaldecodingcomplexity}
\begin{split}
O&_{\emph{\emph{Traditional}}}=\frac{1}{T}\sum^{1}_{l=L}[M_c<M^{L-l+1}]M\\
&\cdot\left([M_c> M^{L-l}]\right.\left.M^{L-l}+[M_c\leq
M^{L-l}]M_c\right),
\end{split}
\end{equation}
\noindent where $[Condition]$ will be 1 when $Condition$ is true, or 0 when $Condition$ is false. In (\ref{eq_normaldecodingcomplexity}), $\frac{1}{T}$ means that the decoding complexity is averaged over the symbol durations; $l=L,\cdots, 1$ means that the decoding process is conducted from $s_L$ to $s_1$; $[M_c<M^{L-l+1}]$ being 1 means that the number of total search paths exceeds $M_c$ and Euclidean metrics need to be calculated.

Note that 1) if $M_c = 1$, traditional QRDM has successive interference cancelation (SIC) and
(\ref{eq_normaldecodingcomplexity}) becomes $O_{\emph{\emph{Traditional}}}=\frac{1}{T}LM$ \emph{(the first point
seems useless)}; 2) if $M_c=M^{L-k}$, traditional QRDM becomes the same complexity as fast decoding \cite{Biglieri} and the decoding complexity is reduced from $\frac{1}{T}M^L$ to $O_{\emph{\emph{Traditional}}}=\frac{1}{T}kMM_c=\frac{1}{T}kM^{L-k+1}$. In this case, only the orthogonality in the upper-left sub-block of the block-orthogonal structure is exploited. In the following, we will propose a simplified decoding that exploits the orthogonality in all the sub-blocks of block-orthogonal structure to reduce $M_c$
to $M^{eq}_c$ for Euclidean metric calculations, and hence reducing the decoding complexity (\ref{eq_normaldecodingcomplexity}) further, but without performance loss.

\subsection{Simplified QRDM for Block-Orthogonal Structure} \label{section_comp_bosmc_qrdm}
\begin{figure}[!t]
\centering
\includegraphics[width=2.6in]{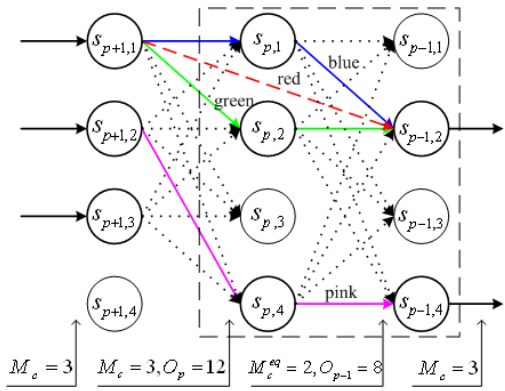}
\caption{Simplified QRDM trellis diagram ($M=4$).} \label{fig_qrdv}
\end{figure}

As denoted in the dashed box of Fig. \ref{fig_qrdv}, we assume that 2 real information symbols $\{s_{p},s_{p-1}\}$ drawn from a signal constellation with $M=4$ are in a sub-block of an STC with block-orthogonal structure $(\Gamma, k,1)~(k\geq2)$ and this sub-block is not a first-decoded block. Since $s_p$ and $s_{p-1}$ are independent, the Euclidean metric calculations for $s_p$ and $s_{p-1}$ can be separated. Under QRDM with $M_c=3$, the Euclidean metric calculation number for $s_p$ is $O_p=M_cM=12$. Without loss of generality, the surviving candidates for $s_{p}$ may be $s_{p,1}, s_{p,2}$ and $s_{p,4}$ (as shown in Fig. \ref{fig_qrdv}) based on the updated accumulated Euclidean distance.

Next we calculate the Euclidean metrics for $s_{p-1}$. Since $s_p$ and $s_{p-1}$ are orthogonal to each other, $s_p$ will not affect the Euclidean metric calculations for $s_{p-1}$. Hence the two Euclidean metrics for $s_{p-1}$ along path
\begin{equation*}
(\cdots \rightarrow s_{p+1,1}\rightarrow s_{p,1} \rightarrow
s_{p-1,j}):\emph{\emph{ blue line in Fig. \ref{fig_qrdv}}}~~~~~~~~
\end{equation*}
and path
\begin{equation*}
(\cdots \rightarrow s_{p+1,1}\rightarrow s_{p,2} \rightarrow
s_{p-1,j}):\emph{\emph{ green line in Fig. \ref{fig_qrdv}}}~~~~~~~
\end{equation*}
are the same, which is equal to the Euclidean metric along the virtual path
(for Euclidean metric calculation only)
\begin{equation*}
(\cdots \rightarrow s_{p+1,1}\rightarrow s_{p-1,j}): \emph{\emph{
red dashed line in Fig. \ref{fig_qrdv}}}
\end{equation*}
with $j=1,2,3$ and $4$. Then the number of Euclidean metric calculation for $s_{p-1}$ will be reduced from $M_{c}M=12$ to
$O_{p-1}=M^{eq}_{c}M=8$ where $M^{eq}_{c}=2$ is the equivalent surviving path
number of $\{\cdots s_{p+2},s_{p+1},s_{p}\}$ for
$s_{p-1}$, or the surviving path number of $\{\cdots
s_{p+2},s_{p+1}\}$ for $s_{p-1}$. Note that the red virtual path does not exist under traditional QRDM without considering block-orthogonal structure. Thus, under proposed simplified QRDM considering block-orthogonal structure, the surviving paths for
Euclidean metric calculations are the $M^{eq}_{c}$ reserved paths of
these symbols $\{\cdots s_{p+2},s_{p+1}\}$ decoded in previous blocks, not the $M_{c}$ reserved paths of all these symbols $\{\cdots s_{p+2},s_{p+1},s_{p}\}$ decoded previously. Note that with the use of virtual path, we can reduce the Euclidean metric calculation number, but will not reduce the surviving path number. Hence, the decoding complexity reduction will not cause any loss in BER performance at all. Without loss of generality, after updating the accumulated Euclidean distance, $M_c=3$ surviving paths in this stage may be
\begin{equation*}
\begin{split}
&(1)~\cdots\rightarrow s_{p+1,1}\rightarrow s_{p,1}\rightarrow
s_{p-1,2}
:\emph{\emph{ blue line in Fig. \ref{fig_qrdv}}}\\
&(2)~\cdots\rightarrow s_{p+1,1}\rightarrow s_{p,2}\rightarrow
s_{p-1,2}
:\emph{\emph{ green line in Fig. \ref{fig_qrdv}}}\\
&(3)~\cdots\rightarrow s_{p+1,2}\rightarrow s_{p,4}\rightarrow
s_{p-1,4}
:\emph{\emph{ pink line in Fig. \ref{fig_qrdv}}}\\
\end{split}
\end{equation*}

Suppose that $\{s_{p},s_{p-1},\cdots,s_{p-k+1}\}$ are in the
same sub-block of block-orthogonal structure $(\Gamma,k,1)$ and their
equivalent surviving path numbers are denoted as
$M^{eq}_{c,p},M^{eq}_{c,p-1},\cdots,M^{eq}_{c,p-k+1}$, respectively. Then, instead of (\ref{eq_normaldecodingcomplexity}), we can rewrite the decoding complexity under proposed simplified QRDM as:
\begin{equation} \label{eq_bosmcdecodingcomplexity}
\begin{split}
O&_{\emph{\emph{Simplified}}}=\frac{1}{T}\sum^{1}_{l=L}{[M_c<M^{L-l+1}]M}\\
&\cdot\left([M_c> M^{L-l}]\right.\left.M^{L-l}+[M_c\leq M^{L-l}]M^{eq}_{c,l}\right).
\end{split}
\end{equation}

It is easy to see that $M_c=M^{eq}_{c,p}\geqslant M^{eq}_{c,p-1}
\geqslant\cdots \geqslant M^{eq}_{c,p-k+1}$. Hence, the decoding
complexity of an STC with block-orthogonal structure of $k\geq 2$
under proposed simplified QRDM is lower than that under traditional
QRDM.

\subsubsection{Decoding Complexity Reduction Bound}
Considering a sub-block $\{s_{p},s_{p-1},\cdots,s_{p-k+1}\}$ of block-orthogonal structure $(\Gamma,k,1)$, it is easy to see that $M^{eq}_{c,p-i}$ achieves the minimum value of $\frac{M_{c}}{M^{i}}$ $(i=0,\cdots,k-1)$ when each node in the reserved decoding search tree has $M$ children and hence as few parent nodes as possible are reserved. The minimum simplified decoding complexity of this sub-block can be shown to be (assume that $M_c\leq M^{L-p}$)
\begin{equation*}
\begin{split}
O_{\emph{\emph{part Simplified,min}}}
&=\sum^{p-k+1}_{i=p}{MM^{eq}_{c,i}}=\sum^{k-1}_{i=0}{\frac{MM_c}{M^{i}}}\\
&=\frac{M^{k}-1}{M^{k}-M^{k-1}}MM_c\approx\frac{M}{M-1}MM_c.
\end{split}
\end{equation*}
Compared with the traditional per-sub-block decoding complexity $kMM_c$ ( when $M_c\leq M^{L-p}$), the
simplified decoding complexity of this sub-block can even be reduced to $\frac{M}{k(M-1)}$. Note that this result is available to all non-first-decoded sub-blocks. For the block-orthogonal structure $(\Gamma,k,\gamma)$ with $\gamma>1$, the $\gamma$ information symbols corresponding to an upper-triangular matrix $\textbf{U}$ should be viewed as a unit drawn from a constellation of size $M^\gamma$, instead of $M$. Hence we can make the following remark:
\begin{remark} \label{remark1}
\emph{Compared with the traditional decoding, the maximum amount of decoding complexity reduction of block-orthogonal structure $(\Gamma,k,\gamma)$ with simplified decoding is $\frac{M^{\gamma }}{k(M^{\gamma }-1)}$ approximately, which is a decreasing function of $k$ and $M^\gamma$.}
\end{remark}

\subsubsection{Decoding Complexity Reduction Example}
From (\ref{eq_bosmcdecodingcomplexity}), we can see that the
decoding complexity is mainly determined by the equivalent surviving
path number $M^{eq}_{c,l}$. Since the actual $M^{eq}_{c,l}$ value
can only be estimated experimentally, simulations are conducted in a
4$\times$2 MIMO system where the D-STTD code with block-orthogonal
structure $(2,4,1)$ is applied, and $M^{eq}_{c,l}$ of
$s_l(l=4,\cdots,1)$ in the non-first-decoded sub-block are
enumerated. In the simulations, the communication channel is assumed
to be quasi-static Rayleigh fading and the channel state information
is perfectly known at the receiver.

\begin{figure}[!t]
\centering
\includegraphics[width=3.2in]{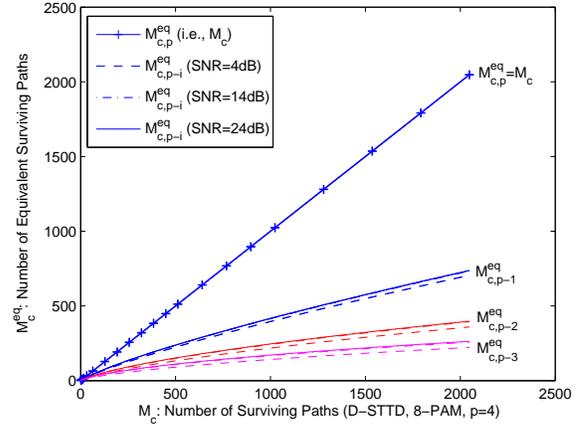}
\caption{$M^{eq}_{c,p-i}~(i=0,\cdots,k-1,~p=4,~k=4)$ for the D-STTD code where each real information symbol is drawn from 8-PAM.}
\label{rdsttd64qamMvmultidB}
\end{figure}

\begin{figure}[!t]
\centering
\includegraphics[width=3.2in]{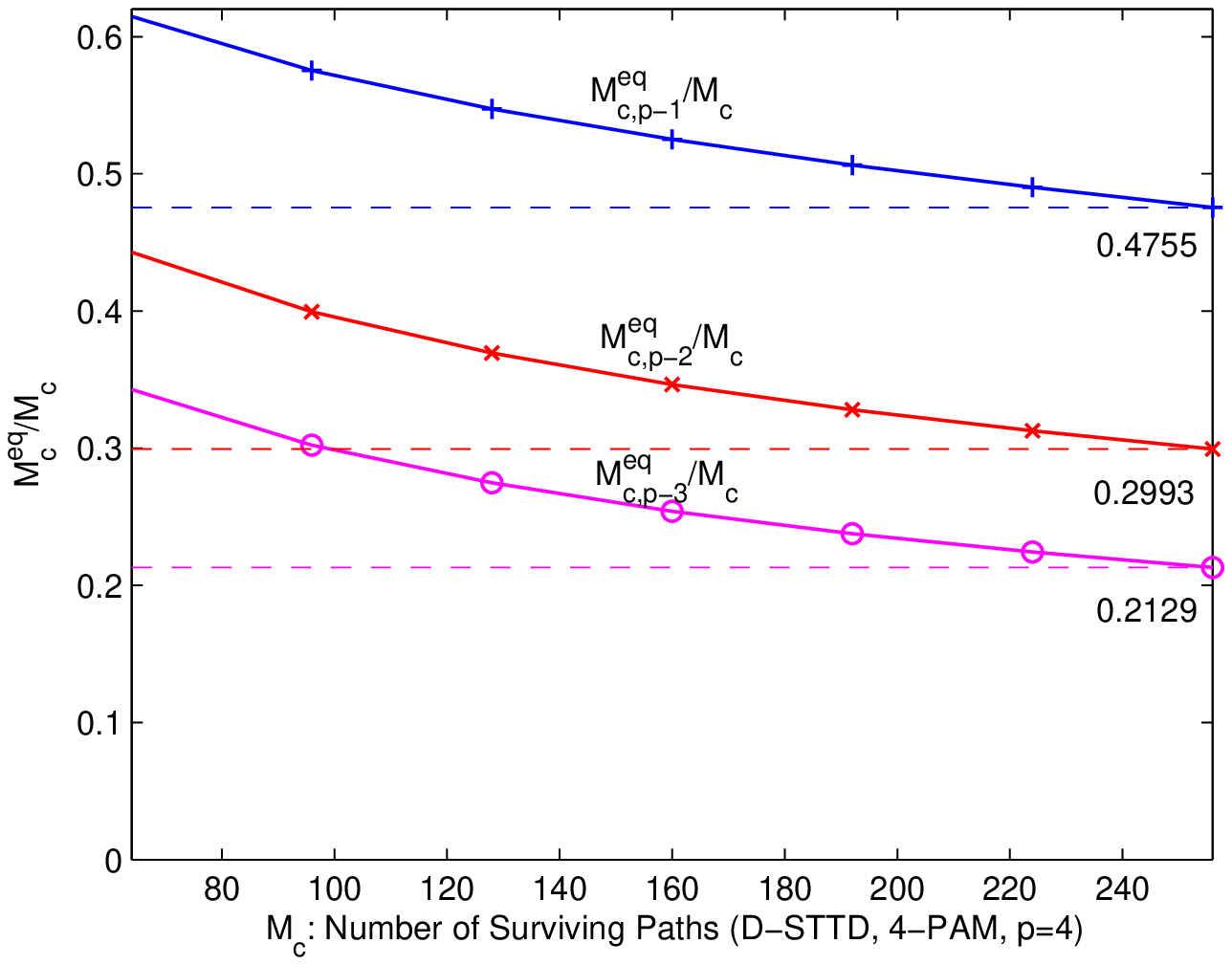}
\caption{$M^{eq}_{c,p-i}/M_c~(i=1,\cdots,k-1,~p=4,~k=4)$ for the D-STTD code where each real information symbol is drawn from 4-PAM.} \label{rdsttd16qamcomp}
\end{figure}

\begin{figure}[!t]
\centering
\includegraphics[width=3.2in]{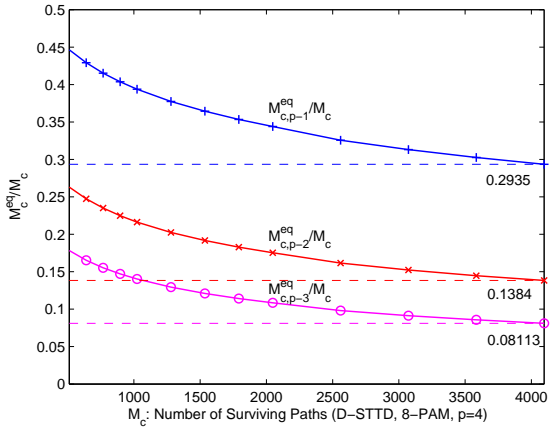}
\caption{$M^{eq}_{c,p-i}/M_c~(i=1,\cdots,k-1,~p=4,~k=4)$ for the D-STTD code where each real information symbol is drawn from 8-PAM.} \label{rdsttd64qamcomp}
\end{figure}

Experiment I: Assume that each real information symbol in the D-STTD code is drawn from 8-PAM, the equivalent surviving path numbers under different SNR values 4dB, 14dB, 24dB are shown in Fig. \ref{rdsttd64qamMvmultidB}. We can see that all the equivalent surviving path numbers $M^{eq}_{c,p-i}~(i=1,\cdots,k-1,~p=4,~k=4)$ are much smaller than $M_c$.

Experiment II: Assuming that each real information symbol in the D-STTD code is drawn from 4-PAM and 8-PAM in 2 simulations, respectively. The complexity reduction results of $M^{eq}_c/M_c$ are shown in Fig. \ref{rdsttd16qamcomp} and Fig. \ref{rdsttd64qamcomp}, where we can see that $M^{eq}_{c,p-i}/M_c$ is far smaller than 1, and it decreases with increasing $i$ $(i=0,\cdots,k-1,~k=4)$ and $M$ (i.e., $M^\gamma$).

\begin{remark}\label{remark4}
\emph{In the proposed simplified QRDM, the decoding complexity (\ref{eq_bosmcdecodingcomplexity}) decreases with increasing $k$ and $M^\gamma$, and the maximum complexity reduction order concurs with Remark \ref{remark1}.}
\end{remark}

\subsection{Decoding Complexity Comparisons}\label{seccomparison}
\begin{figure}[!t]
\centering
\includegraphics[width=3.6in]{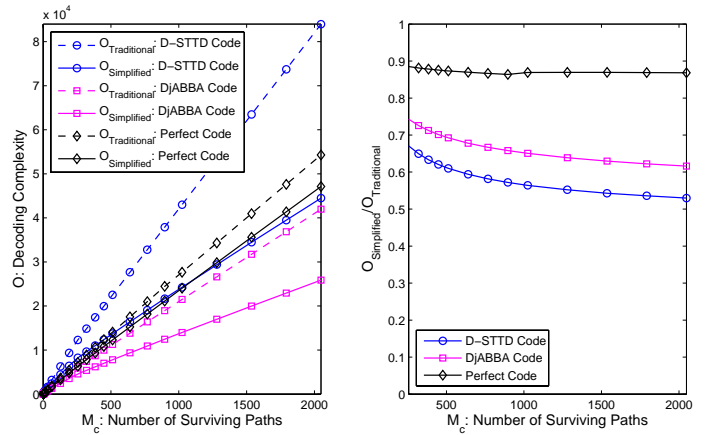}
\caption{(a) Traditional and proposed simplified decoding complexity and (b) Ratio between traditional and proposed simplified decoding complexity.}
\label{fig_decodingcomp_comparison3}
\end{figure}

Under traditional and proposed simplified QRDM\textquoteright s, the decoding complexities of the D-STTD code \cite{Texas_dsttd}, the DjABBA code \cite{Hottinen} and the Perfect code \cite{Oggier} with block-orthogonal structures ($2,4,1$), ($4,2,2$) and ($16,2,1$) respectively in a 4$\times$4 MIMO system are compared in Fig. \ref{fig_decodingcomp_comparison3}, where each real information symbol is drawn from 16-PAM, 4-PAM and 4-PAM, respectively. We emphasize that all codes will have exactly the same BER performance under both QRDM schemes because the proposed Simplified QRDM only reduces the number of Euclidean metric calculations but not the surviving path number (as explained earlier in Section IV-B). From Fig. \ref{fig_decodingcomp_comparison3}, we can see that the decoding complexity under proposed simplified QRDM can be reduced drastically. In particular, the complexity reduction for the D-STTD code is nearly 50\%.

Moreover, the simulation also shows that: 1) the D-STTD code achieves more decoding complexity reduction than the DjABBA code. That is because compared to the DjABBA code with $k=2$, the D-STTD code has larger $k=4$ (both have the same $M = 16$); 2) the DjABBA code achieves more decoding complexity reduction than Perfect code because the DjABBA code has a larger $M=4^2=16$ than the Perfect code which has $M=4^1=4$ (both have the same $k = 2$). Both the observations concur with Remark \ref{remark4}.

Note that although the proposed Simplified QRDM achieves a lower decoding complexity, its BER performance remains the same as the traditional QRDM because the surviving path number of both schemes remain the same (recall explanation in Section IV-B). Hence, BER comparisons under traditional and proposed simplified QRDM\textquoteright s are unnecessary and omitted.

\section{New BOSTC Construction}\label{sec_bostcconstruct}

Although we have shown that many existing high-rate STCs have some block-orthogonal structure, there are new open problems:

1) The conventional approaches to high-rate code design tend to focus on the error rate performance criteria and always ignore decoding complexity, hence they may not achieve the best performance-complexity trade off. We can see that for most existing codes in Table \ref{table_existingcode}, $k$ is 2 hence the decoding complexity reduction under proposed simplified QRDM is limited. Furthermore, the Perfect code with 3 and 6 transmit antennas can not benefit from simplified QRDM decoding due to $k=1$;

2) Many existing BOSTC have low scalability. For example, the maximum code rate of DjABBA code is 2.

Therefore in this section, we will construct new BOSTC's which better exploit the
block-orthogonal property and are more scalable.

\subsection{Construction}
To reduce the decoding complexity, the BOSTC should be designed for
large $k$. Two such systematic construction rules are presented
here.

\subsubsection{Construction I with Rate-1 Seed Code}
Select a rate-1 space-time code $\textbf{X}_{o,T\times N_t}$ with high $k$ to be the seed code, and a full-rank matrix $\textbf{M}_{N_t\times N_t}=\left[\textbf{m}_1~\textbf{m}_2~\cdots~\textbf{m}_{N_t}\right]$ to be the extension matrix, then a rate-$N_t$ (full-rate) BOSTC can be constructed as
\begin{equation}\label{bostc4txI}
\textbf{X}_{\textbf{\textbf{I}},N_t}=\sum^{N_t}_{i=1}{\textbf{X}_{o,i}\cdot
\emph{\emph{diag}}(\textbf{m}_i)}
\end{equation}
\noindent where $\textbf{X}_{o,i}$ is the $\textbf{X}_{o}$ with different sets of information symbols. It is easy to prove that the decoding of $\textbf{X}_{\emph{\emph{I}},N_t}$ is not rank deficient if there is at most a complex information symbol at each space-time position of the seed code $\textbf{X}_{o,T\times N_t}$.

\subsubsection{Construction II with Rate-1/2 Seed Code}
Select a rate-1/2 space-time code $\textbf{X}_{o,T\times N_t}$ with high $k$ to be the seed code, and a matrix $\textbf{M}_{N_t\times 2N_t}=\left[\textbf{m}_1~\textbf{m}_2~\cdots~\textbf{m}_{2N_t}\right]$ with full-rank
$\tqbinom{\textbf{M}^R}{\textbf{M}^I}$ be the extension matrix, then a rate-$N_t$ BOSTC can be constructed as
\begin{equation}\label{bostc4txII}
\textbf{X}_{\emph{\emph{II}},N_t}=\sum^{2N_t}_{i=1}{\textbf{X}_{o,i}\cdot
\emph{\emph{diag}}(\textbf{m}_i)}
\end{equation}
\noindent where $\textbf{X}_{o,i}$ is the $\textbf{X}_{o}$ with different sets
of information symbols. It is easy to prove that the decoding of $\textbf{X}_{\emph{\emph{II}},N_t}$ is not rank deficient if there is at most a real information symbol at each space-time position of the seed code $\textbf{X}_{o,T\times N_t}$.

\subsection{Examples}
BOSTC examples with $k=4$ and $k=8$ are presented in the following. To our knowledge, the $k=8$ code has the largest $k$ value ever reported.

First we review Hadamard matrix. A complete set of $2^m$ Walsh functions of order $m$ gives a Hadamard matrix $\textbf{M}_{2^m}$ \cite{Thompson} as follows:
\begin{equation}\label{H}
\begin{split}
\textbf{M}_2&=\left[
            \begin{array}{cc}
              1 & 1 \\
              1 & -1 \\
            \end{array}
          \right],~
\textbf{M}_4=\left[
            \begin{array}{cc}
              \textbf{M}_2 & \textbf{M}_2 \\
              \textbf{M}_2 & -\textbf{M}_2 \\
            \end{array}
          \right],\cdots,\\
\textbf{M}_{\infty}&=\cdots
\end{split}
\end{equation}

Denote $\textbf{M}_{N}$ as a square sub-matrix of $\textbf{M}_{\infty}$ in (\ref{H}) formed by the first $N$ columns and the first $N$ rows. For example,
\begin{equation*}
\textbf{M}_{3}=\left[\begin{array}{ccccccccccccc}
              1 & 1   &1\\
              1 & -1   &1\\
              1 & 1   &-1\\
            \end{array}
          \right].
\end{equation*}

\vspace{0.06in}
\noindent \textbf{Example 1} (fixed dimension): $(8,4,1)$-BOSTC for 4 transmit antennas

Let the seed code be the rate-1 jABBA code:
\begin{equation}\label{boSTC4txIm}
\begin{split}
\textbf{X}_{o}=\left[
\begin{array}{cccccccc}
    s_1+js_2       &   s_3+js_4       & js_5-s_6      &   js_7-s_8 \\
    -s_3+js_4      &   s_1-js_2       & -js_7-s_8     &   js_5+s_6\\
    s_5+js_6       &   s_7+js_8       & s_1+js_2      &   s_3+js_4 \\
    -s_7+js_8      &   s_5-js_6       & -s_3+js_4     &   s_1-js_2\\
\end{array}
\right]
\end{split}
\end{equation}

Then a rate-4 STC $\textbf{X}_{\emph{\emph{I}},4}$ for 4 transmit antennas can be constructed as
\begin{equation} \label{bostc4txI1}
\textbf{X}_{\emph{\emph{I}},4}=\sum^{4}_{i=1}{\textbf{X}_{o,i}\cdot
\emph{\emph{diag}}(\textbf{m}_i)}
\end{equation}
\noindent where $\textbf{X}_{o,i}$ is the $\textbf{X}_{o}$ with different sets of information symbols $\{s_{1,i},\cdots, s_{8,i}\}$ and $\textbf{m}_i$ is the $i$th column of Hadamard matrix $\textbf{M}_{4}$ as shown in (\ref{hadamard4x4}).

\begin{equation}\label{hadamard4x4}
\begin{split}
\textbf{M}_{4}=\left[
\begin{array}{cccccccc}
    1      &   1      & 1     &  1 \\
    1      &   -1     & 1     &  -1  \\
    1      &  1       & -1    &   -1  \\
    1      &   -1     & -1    &  1 \\
\end{array}
\right]
\end{split}
\end{equation}

Following Theorem \ref{th_bostbc_b}, $\textbf{X}_{\emph{\emph{I}},4}$ can be verified \cite{codeexample_matlab} to have block-orthogonal structure $(8,4,1)$, where $\{s_{1,i},\cdots,s_{4,i}\}$ and $\{s_{5,i},\cdots, s_{8,i}\}$ are in the $(2i-1)$th and $(2i)$th sub-blocks, respectively. Moreover, the block-orthogonal structure is maintained even if any of these sub-blocks are removed, and hence $\textbf{X}_{\emph{\emph{I}},4}$ can be a $(\Gamma,4,1)$-BOSTC of code rate
$\Gamma/2~(\Gamma=1,2,\cdots,8)$ with $(8-\Gamma)$ sub-blocks removed.

\vspace{0.06in}
\noindent \textbf{Example 2}(scalable dimension): $(2^{m+n-1},4,2^{m-n})$-BOSTC for $2^m$ transmit antennas $(m\geq1~\emph{\emph{integer}},~n\in [1,~m])$

Let $\textbf{C}_{l,1,1}(l=1,2,3\emph{\emph{ and }}4)$ be the dispersion matrices of Alamouti code \cite{Alamouti} with $j^2=-1$:
\begin{equation} \label{alamouti}
\begin{split}
\left[
\begin{array}{cccc}
    1      &   0     \\
    0      &   1    \\
\end{array}
\right],\left[
\begin{array}{cccc}
    j      &   0     \\
    0      &   -j    \\
\end{array}
\right],\left[
\begin{array}{cccc}
    0      &   1     \\
    -1     &   0    \\
\end{array}
\right],\left[
\begin{array}{cccc}
    0      &   j     \\
    j      &   0    \\
\end{array}
\right].
\end{split}
\end{equation}

\noindent Then the dispersion matrices of a rate-1 STC for $2^m(m>1\emph{\emph{ integer}})$ transmit antennas can be presented as:
\begin{equation} \label{original_4gp}
\begin{split}
\textbf{C}_{l,2k-1,m}&=\left[
\begin{array}{cccc}
    \textbf{C}_{l,k,m-1}      &   \textbf{0}     \\
    \textbf{0 }               &   \textbf{C}_{l,k,m-1}    \\
\end{array}
\right],\\
\textbf{C}_{l,2k,m}&=\left[
\begin{array}{cccc}
    \textbf{0}              & \textbf{C}_{l,k,m-1}       \\
    \textbf{C}_{l,k,m-1}    &   \textbf{0}    \\
\end{array}
\right]
\end{split}
\end{equation}
\noindent where $k=1,\cdots,2^{m-2}$, $l=1,2,3\emph{\emph{ and }}4$.

\begin{figure}[!t]
\centering
\includegraphics[width=2.8in]{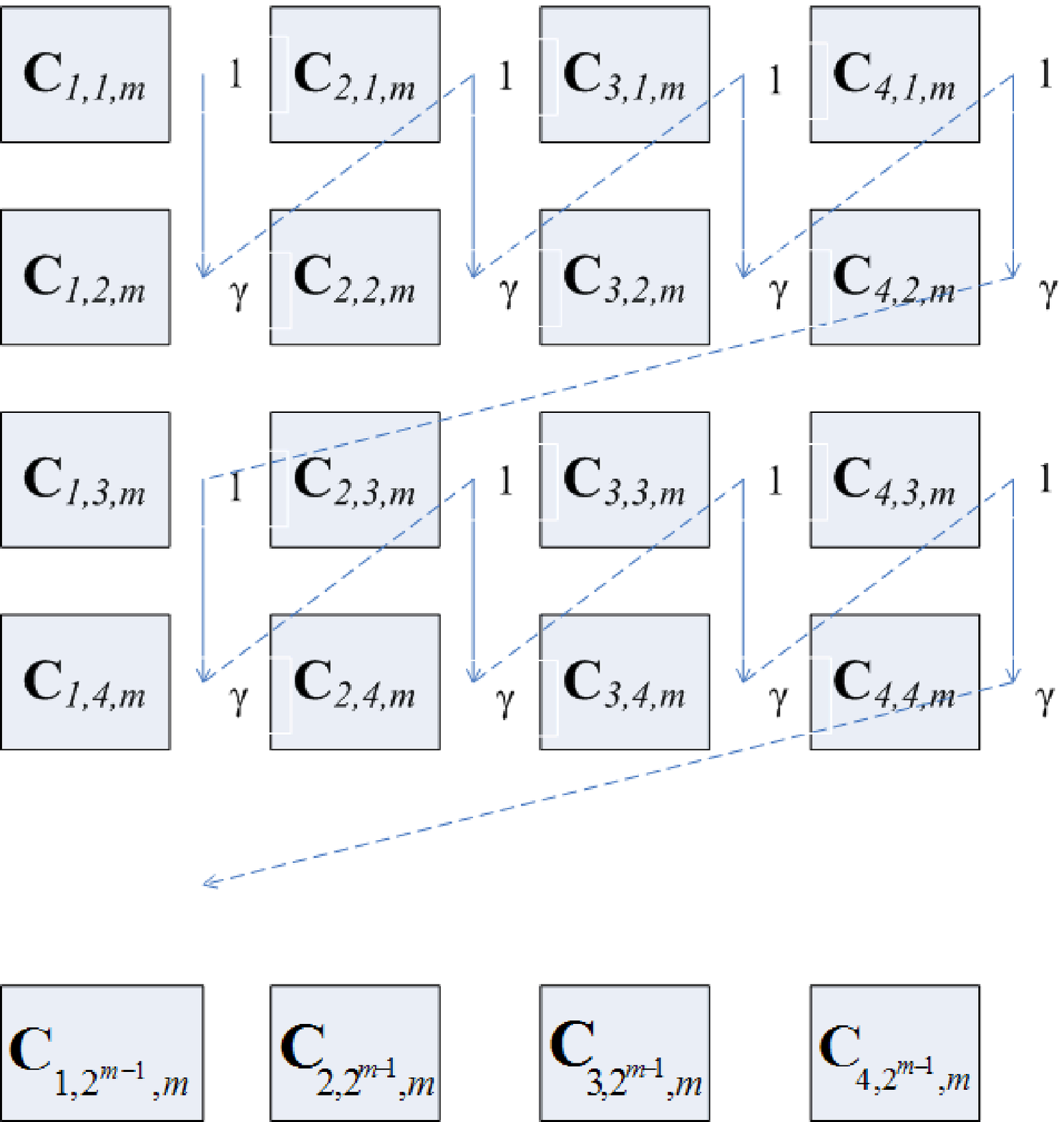}
\caption{Order of picking dispersion matrices ($\gamma=2^{m-n}$, $n\in [1,~m]$, $m\geq1$) in {Example 2}. Here $\gamma=2$ for illustration purpose.}
\label{fig_bostc_example}
\end{figure}

The rate-1 STC with dispersion matrices in (\ref{alamouti}) or (\ref{original_4gp}) with the picking order shown in Fig \ref{fig_bostc_example} is denoted as $\textbf{X}_{o}$. Let the seed code be $\textbf{X}_{o}$ and the extension matrix $\textbf{M}$ be the Hadamard matrix of size $2^m\times 2^m$, a rate-$2^m$ STC $\textbf{X}_{\emph{\emph{I}},2^m}$ can be constructed following Construction I:
\begin{equation}\label{bostc4txI2}
\textbf{X}_{\emph{\emph{I}},2^m}=\sum^{2^m}_{i=1}{\textbf{X}_{o,i}\cdot
\emph{\emph{diag}}(\textbf{m}_i)}
\end{equation}
\noindent where $\textbf{X}_{o,i}$ is the rate-1 STC $\textbf{X}_{o}$ with different sets of information symbols and
$\textbf{m}_i$ is the $i$th column of Hadamard matrix $\textbf{M}_{2^m}$.

Following Theorem \ref{th_bostbc_b}, $\textbf{X}_{\emph{\emph{I}},2^m}$ can be verified \cite{codeexample_matlab} to have block-orthogonal structure $(2^{m+n-1},4,2^{m-n})$ with $n\in [1,m]$, where $\{s_{l,(k-1)\gamma+1,m,i},\cdots,s_{l, k\gamma ,m,i}\}$ corresponds to $\textbf{U}_{p,l}$ ($l=1,2,3\emph{\emph{ and }}4$) in the $p$th sub-block ($k=1,\cdots,2^{n-1}$, $i=1,\cdots,2^m$, $p=2^{n-1}(i-1)+k$). Moreover, the block-orthogonal structure is maintained even if some sub-blocks are removed, hence $\textbf{X}_{\emph{\emph{I}},2^m}$ can be a $(\Gamma,4,2^{m-n})$-BOSTC of code rate $2^{1-n}\Gamma~(\Gamma=1, \cdots,2^{m+n-1})$ with $(2^{m+n-1}-\Gamma)$ sub-blocks removed.

Using the rate-$1/2$ real orthogonal STC in \cite{Tarokh_ostbc} as the seed codes, BOSTC can be obtained following Construction II as follows.

\vspace{0.06in}
\noindent \textbf{Example 3}: $(10,8,1)$-BOSTC for 5 transmit antennas

Let the seed code be
\begin{equation}\label{Xm_II1081}
\begin{split}
\textbf{X}_{o}=\left[
\begin{array}{ccccccccccccccccccc}
    s_1        &   s_2        & s_3       &   s_4    & s_5        \\
    -s_2       &   s_1        & s_4       &   -s_3   & s_6        \\
    -s_3       &   -s_4       & s_1       &   s_2    & s_7        \\
    -s_4       &   s_3        & -s_2      &   s_1    & s_8        \\
    -s_5       &   -s_6       & -s_7      &   -s_8   & s_1        \\
    -s_6       &   s_5        & -s_8      &   s_7    & -s_2       \\
    -s_7       &   s_8        & s_5       &   -s_6   & -s_3       \\
    -s_8       &   -s_7       & s_6       &   s_5    & -s_4       \\
\end{array}
\right],
\end{split}
\end{equation}
and the extension matrix be
\begin{equation}\label{E_II1081}
\begin{split}
\textbf{M}=\left[
\begin{array}{cccccccccccccccccccccccccccccc}
    -1 & 1 & 1 & 1 & 1 & j & 1 & 1 & 1 & 1  \\
    1 & -1 & 1 & 1 & 1 & 1 & j & 1 & 1 & 1  \\
    1 & 1 & -1 & 1 & 1 & 1 & 1 & j & 1 & 1  \\
    1 & 1 & 1 & -1 & 1 & 1 & 1 & 1 & j & 1  \\
    1 & 1 & 1 & 1 & -1 & 1 & 1 & 1 & 1 & j  \\
\end{array}
\right],
\end{split}
\end{equation}
a rate-5 STC $\textbf{X}_{\emph{\emph{II}},5}$ can be constructed following Construction II:
\begin{equation}\label{bostc5txII1}
\textbf{X}_{\emph{\emph{II}},5}=\sum^{10}_{i=1}{\textbf{X}_{o,i}\cdot
\emph{\emph{diag}}(\textbf{m}_i)}
\end{equation}
\noindent where $\textbf{X}_{o,i}$ is $\textbf{X}_{o}$ in (\ref{Xm_II1081}) with different sets of information symbols and $\textbf{m}_i$ is the $i$th column of $\textbf{M}$ in (\ref{E_II1081}).

Following Theorem \ref{th_bostbc_b}, $\textbf{X}_{\emph{\emph{II}},5}$ can be verified to have block-orthogonal structure $(10,8,1)$, where $\{s_{1,i},\cdots,s_{8,i}\}$ are in the $i$th $(i=1,\cdots,10)$ sub-block. Note that the block-orthogonal structure is maintained even if some sub-blocks are removed, hence $\textbf{X}_{\emph{\emph{II}},5}$ can be a $(\Gamma,8,1)$-BOSTC of code rate $\Gamma/2~(\Gamma=1,2,\cdots,10)$ with $(10-\Gamma)$ sub-blocks removed.

The newly constructed BOSTC are summarized in Table \ref{table_summary_propcode}. Interestingly, the $\textbf{X}_{\emph{\emph{II}},5}$ code found using Construction II has a higher $k$ value (=8) than those found using Construction I. $\textbf{X}_{\emph{\emph{II}},5}$ is also the first ever $k=8$ code.

\begin{table*}[!t]
\begin{center}
\begin{threeparttable}[!b]
\tabcolsep 6mm
\caption{Comparison of BOSTC for $N_t$ Transmit Antennas over $T$ Symbol Durations$^a$.}
\label{table_summary_propcode}
\newcommand{\rb}[1]{\raisebox{2.8ex}[0pt]{#1}}
\newcommand{\rbb}[1]{\raisebox{1.0ex}[0pt]{#1}}
{\small\begin{tabular}{|c|c|c|c|c|c|c|} \hline {\rule[-1mm]{0mm}{5mm}} BOSTC  &  $N_t$ & $T$
& $Rate$ & $\Gamma$ & $k$ & $\mathbbm{k}$
\\\hline\hline
{\rule[-1mm]{0mm}{5mm}}$\textbf{X}_{\emph{\emph{I}},4}$ in (\ref{bostc4txI1}) & $4$ & $4$ & $4$    & $8$ & \textbf{4} & $1$
\\ \hline
{\rule[-1mm]{0mm}{5mm}}$\textbf{X}_{\emph{\emph{I}},2^m}$ in (\ref{bostc4txI2})$^b$ & $2^m$ & $2^m$ & $2^m$    & $2^{m+n-1}$ &
\textbf{4} & $2^{m-n}$
\\ \hline
{\rule[-1mm]{0mm}{5mm}} $\textbf{X}_{\emph{\emph{II}},5}$ in (22)& 5
& 8 & 5    &10   & \textbf{8} & 1
\\ \hline
\end{tabular}
\begin{tablenotes}
\item[$^a$] Assume that each complex information symbol is drawn from a square QAM without constellation rotation, or each real information symbol is drawn from an one-dimension constellation equivalently;
\item[$^b$] $m $ is an integer $\geq 1$, and $n$ is an integer no larger than $m$.
\end{tablenotes}}
\end{threeparttable}
\end{center}
\hrulefill \vspace*{0pt}
\end{table*}

\subsection{Optimization}
To compare with DjABBA code (rate 2) and DSTTD code (rate 2), we will show a rate-2 BOSTC with optimization in the following.

Denoting $\textbf{X}_{\emph{\emph{o}}}$ in (\ref{boSTC4txIm}) as $\textbf{X}_{\emph{\emph{o}}}=\textbf{X}_{\emph{\emph{o}}_1}(s_1,s_2,s_3,s_4)+
\textbf{X}_{\emph{\emph{o}}_2}(s_5,s_6,s_7,s_8)$, a rate-2 full-diversity $(4,4,1)$-BOSTC $\textbf{X}_{\emph{\emph{I,rate-}}2}$ with optimized design coefficients can be presented as
\begin{equation} \label{bostc4txI11}
\textbf{X}_{\emph{\emph{I,rate-}}2}
=\sum^{1}_{i=0}\sum^{2}_{n=1}{\textbf{X}_{\emph{\emph{o}}_n,i+1}\cdot
\emph{\emph{diag}}(\textbf{p}_{2i+n})\cdot\emph{\emph{diag}}(\textbf{m}_{2i+1})}
\end{equation}
\noindent where $\textbf{X}_{\emph{\emph{o}}_n,i+1}$ is the $\textbf{X}_{\emph{\emph{o}}_n}$ with different sets of information symbols, $\textbf{m}_i$ is the $i$th column vector of Hadamard matrix $\textbf{M}_{4}$ and the design coefficient matrix $\textbf{P}$ can be obtained from computer search as
\begin{equation*}
\begin{split}
\textbf{P}=\left[
\begin{array}{cccccccc}
    \textbf{p}_1  \\
    \textbf{p}_2  \\
    \textbf{p}_3  \\
    \textbf{p}_4  \\
\end{array}
\right]
=\left[
\begin{array}{cccccccc}
    1      &   1       & 1      &  1 \\
    1      &   1       & 1      &  1 \\
    e_1      &  e_1     & e_1    &  e_1 \\
    e_1      &  e_1     & e_1    &  e_1 \\
\end{array}
\right],
\end{split}
\end{equation*}
\noindent where $e_1=e^{j0.3218}$.

\vspace{0.0in}

\section{Simulations and Discussions} \label{sec_Simulation}
In this section, we compare the BER performances of the optimized $\textbf{X}_{\emph{\emph{I,rate-}}2}$ in (\ref{bostc4txI11}) with the existing rate-2 codes such as D-STTD code \cite{Texas_dsttd} and DjABBA code \cite{Hottinen} in 4$\times$2 MIMO systems. We consider the DjABBA code optimized in Chapter 9 of \cite{Hottinen}, which is the best known rate-2 code to our knowledge.

In the following simulations, the proposed simplified QRDM as described in Section \ref{section_comp_bosmc_qrdm} is applied as described, and all the rate-2 codes are modulated by 16-QAM (hence 8 bits/channel use). We assume that the channel is quasi-static Rayleigh fading, and the channel state information (CSI) is known at the receiver perfectly.

\subsection{BER Performance against SNR with Given Decoding Complexities}
From Remark \ref{remark4}, we can see that with a given surviving path number in QRDM, The D-STTD code and the proposed $\textbf{X}_{\emph{\emph{I,rate-}}2}$ in (\ref{bostc4txI11}) with $k=4$ can bring more decoding complexity reduction than the DjABBA code with $k=2$. In other words, with a given decoding complexity, the D-STTD code and the proposed $\textbf{X}_{\emph{\emph{I,rate-}}2}$ support larger surviving path numbers than the DjABBA code. As shown in Table III, we simulate 2 cases in Table \ref{table_bostc_cdt} where Case I considers a decoding complexity $O$ of around 180, while Case II allows a higher decoding complexity $O$ of around 620, for all the D-STTD, DjABBA and proposed $\textbf{X}_{\emph{\emph{I,rate-}}2}$ codes. The complexity order is computed using (\ref{eq_bosmcdecodingcomplexity}) .
\begin{center}
\begin{threeparttable}[!t]
\tabcolsep 4 mm
\caption{QRDM Parameters for Rate-2 Codes: Decoding Complexity $O$ and Surviving Path Number $M_c$.} \label{table_bostc_cdt}
\newcommand{\rb}[1]{\raisebox{2.2ex}[0pt]{#1}}
\newcommand{\rbb}[1]{\raisebox{1.0ex}[0pt]{#1}}
{\small\begin{tabular}{|c|c|c|c|c|}
\hline
{\rule[-0mm]{0mm}{3.0mm}}             &  \multicolumn{2}{c|}{Case I} & \multicolumn{2}{c|}{Case II}
\\ \cline{2-5}
{\rule[-0mm]{0mm}{3.0mm}}             & $M_c$&  $O$ & $M_c$ & $O$
\\ \hline \hline
{\rule[-0mm]{0mm}{3.0mm}} D-STTD code  & 20  &189   & 102  & 622
\\ \hline
{\rule[-0mm]{0mm}{3.0mm}} DjABBA code  & 7 & 208    &  28  &627
\\ \hline
{\rule[-0.0mm]{0mm}{3mm}} $\textbf{X}_{\emph{\emph{I,rate-}}2}$ in (\ref{bostc4txI11})
                                 &  16  & 183 &  64   & 614
\\ \hline
\end{tabular}
}
\end{threeparttable}
\end{center}

\begin{figure}[!t]
\centering
\includegraphics[width=3.6in]{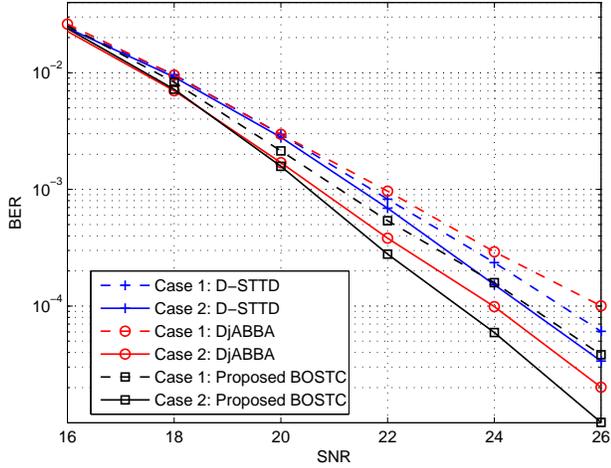}
\caption{BER against SNR with comparable decoding complexities in 4$\times$2 MIMO systems with 8 bits/channel use.}
\label{fig_ber_snr_4rx_8bps}
\end{figure}

The BER curves against SNR are plotted in Fig. \ref{fig_ber_snr_4rx_8bps}. We can see that with similar or slightly
lower decoding complexity(Table \ref{table_bostc_cdt}), $\textbf{X}_{\emph{\emph{I,rate-}}2}$ proposed in (\ref{bostc4txI11}) outperforms both the D-STTD code and the DjABBA code. This is because $\textbf{X}_{\emph{\emph{I,rate-}}2}$ has higher diversity than the D-STTD code, and supports larger surviving
path number than the DjABBA code(see Table \ref{table_bostc_cdt}).

\subsection{BER Performance against Decoding Complexity with Given SNR Value}
From Fig. \ref{fig_ber_snr_4rx_8bps}, we can see that the BER performance of STC decoded using QRDM decoder is a function of the decoding complexity. Interestingly, this function is non-linear. For instance, with similar decoding complexities, the D-STTD code performs better than the DjABBA code in Case I, but worse than the DjABBA code in Case II. Hence in this subsection we will study the relationship between BER performance and decoding complexity under a given SNR value.

\begin{figure}[!t]
\centering
\includegraphics[width=3.6in]{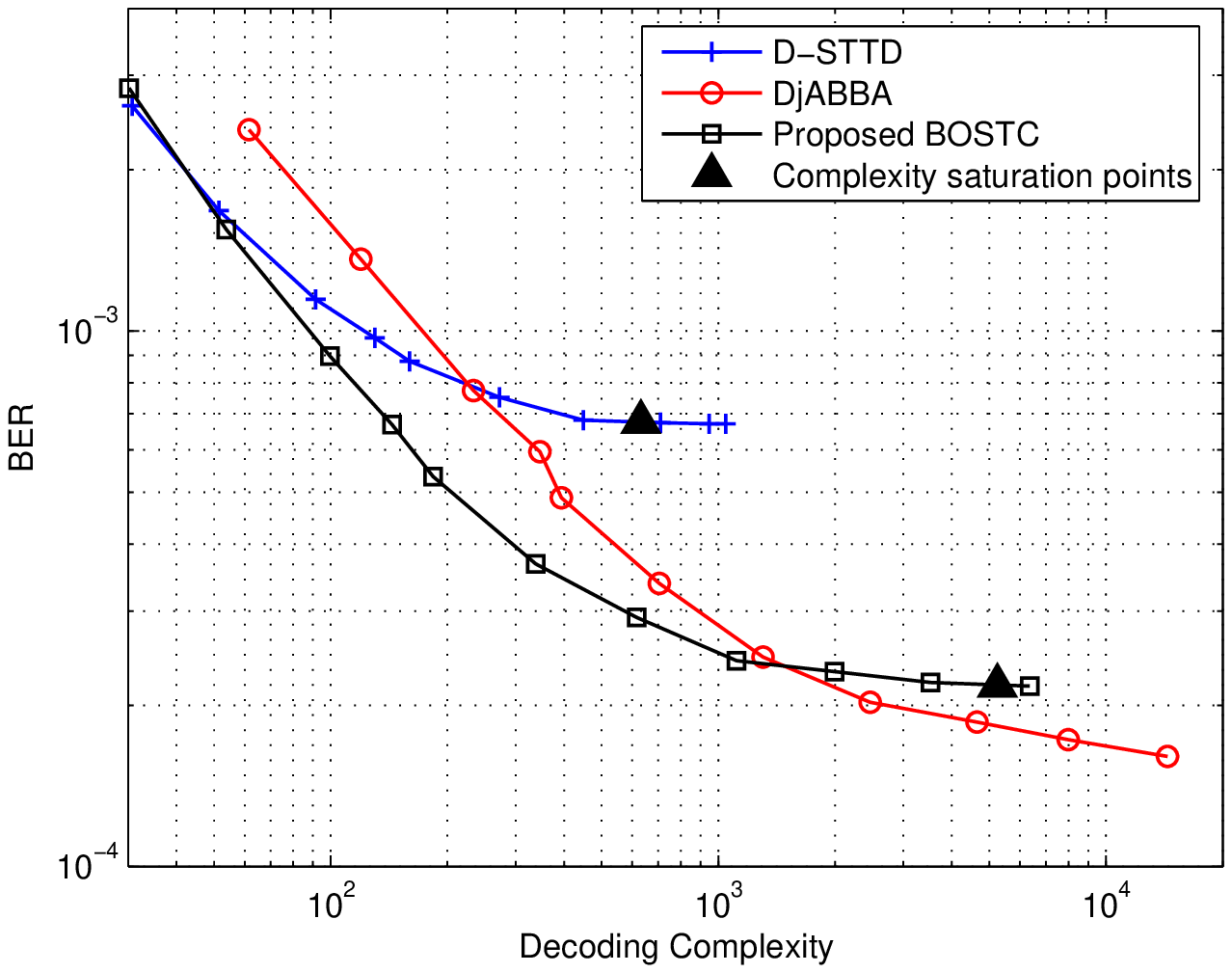}
\caption{BER curves against decoding complexity with a given SNR = 22dB in 4$\times$2 MIMO systems with 8 bits/channel use.}
\label{fig_ber_cdt_4rx_8bps}
\end{figure}

\begin{figure}[!t]
\centering
\includegraphics[width=3.6in]{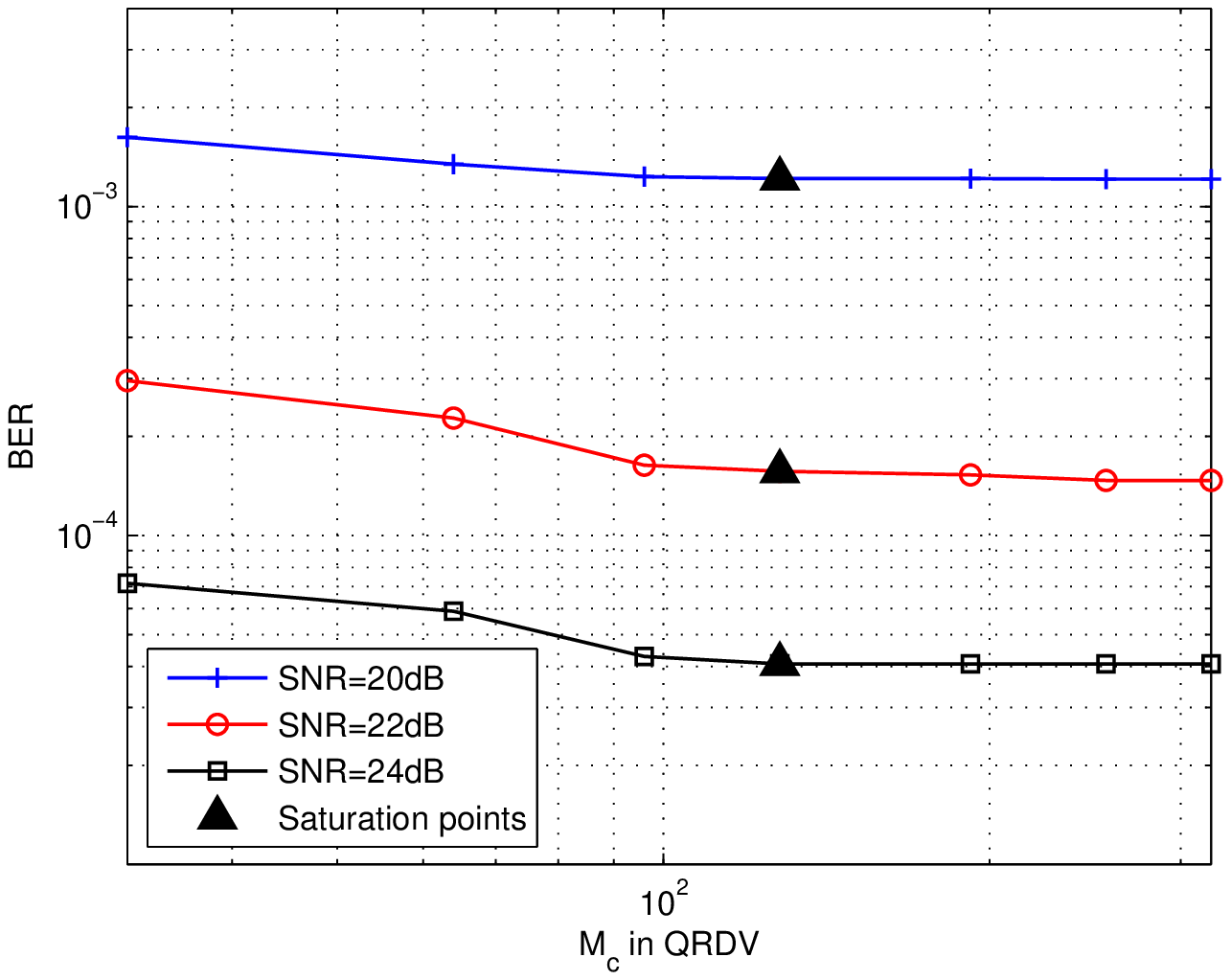}
\caption{BER curves against decoding complexity with a given SNR = 22dB in 4$\times$2 MIMO systems with 8 bits/channel use.}
\label{fig_saturation}
\end{figure}

The BER curves against decoding complexity with SNR = 22 dB are plotted in Fig. \ref{fig_ber_cdt_4rx_8bps}. We can see that 1) at different decoding complexity level, the best performance is achieved by different codes. $\textbf{X}_{\emph{\emph{I,rate-}}2}$ performs the best for most parts of the decoding complexity range, and specifically when the decoding complexity order is lower than $10^3$. Therefore, the proposed BOSTC is a better choice for systems with limited computational power; 2) when the BER curves become flat, the QRDM performance approaches the ML decoding performance, although the practical decoding complexity is far lower than the ML decoding complexity. We call such minimum practical decoding complexity for ML decoding performance the \textbf{complexity saturation
point} and denote it as \textquoteleft \textquoteleft $\blacktriangle$\textquoteright\textquoteright in Fig. \ref{fig_ber_cdt_4rx_8bps} and Fig. \ref{fig_saturation}.

\subsection{Complexity Saturation Point}
From Fig. \ref{fig_ber_cdt_4rx_8bps}, we can see that the codes can achieve near ML decoding performances with a much lower practical decoding complexity (i.e., complexity saturation point) than full ML decoding complexity. When the practical decoding complexity exceeds the complexity saturation point, the improvement on BER performance is trivial. This is a desirable property in high-rate MIMO communication systems.

In Fig. \ref{fig_ber_cdt_4rx_8bps}, the complexity saturation points are obtained with a given SNR = 22 dB. To verify the stability of a code's complexity saturation point, the BER curves of the proposed BOSTC $\textbf{X}_{\emph{\emph{I,rate-}}2}$ with different SNR are plotted in Fig. \ref{fig_saturation}. We can see that the complexity saturation points are almost the same, at about $M_c=128$. This is clearly desirable too.

\section{Conclusions}\label{sec_conclusion}
In this paper, we introduce a new code property, called block-orthogonal property, for space-time codes (STC), and propose a new Simplified QRDM decoder to achieve significant decoding complexity reduction over the traditional breadth-first-search QRDM decoder for many well known high-rate STCs such as the D-STTD, DjABBA and Perfect codes. We prove that the proposed Simplified QRDM has absolutely no performance loss over the traditional QRDM, because the Simplified QRDM reduces only the number of Euclidean metric calculations but not the surviving path number. We also
derive the maximum achievable complexity reduction in terms of the block-orthogonal parameters. To further exploit the block-orthogonal property, we construct new BOSTC with better complexity reduction advantage, and we show how to optimize them for full diversity and maximum coding gain without affecting the block orthogonal code structure. The proposed BOSTC construction rules are scalable, and they support arbitrary number of transmit antennas. Simulations of
BER against SNR and against decoding complexity show that the proposed BOSTC outperforms the best known rate-2 STC under almost all scenarios (except at full ML decoding complexity level), and it requires a QRDM complexity level much lower than the full ML decoding complexity level to achieve near-ML decoding performance.

Finally, we remark that the decoding complexity reduction principle of block-orthogonal code structure presented in both \cite{Ren_bostr} and this paper is applicable to both breadth-first search and depth-first search decoders. Hence, many benefits seen in this paper can also be expected for sphere decoding \cite{Damen}.

\appendices
\section{} \label{proof_th_bostbc_a}

Following the signal model (\ref{eq_r_Hs}), the equivalent
channel matrix in an $N_t\times N_r$ MIMO system with the channel matrix $\tilde{\textbf{H}}_{N_t\times {N_r}} =
[\tilde{\textbf{h}}_1~\tilde{\textbf{h}}_2~\cdots~\tilde{\textbf{h}}_{N_r}]$ is
\begin{equation*}
\begin{split}
&\textbf{H}_{2TN_r\times {L}}
=\left[\textbf{H}_1~\textbf{H}_2\right]
=\left[\textbf{h}_1~\cdots~\textbf{h}_{k}~\textbf{h}_{k+1}~\cdots~\textbf{h}_{2k}\right]
\\&~~~~~~~~~~=\left[
\begin{array}{cccccccc}
    \mathscr{A}_1\bar{\textbf{h}} &\cdots  &\mathscr{A}_k\bar{\textbf{h}}
    &\mathscr{B}_{1}\bar{\textbf{h}} &\cdots  &\mathscr{B}_{k}\bar{\textbf{h}}
\end{array}
\right]
\end{split}
\end{equation*}
\noindent where
\begin{equation*}
\begin{split}
&\bar{\textbf{h}}=\left[
\begin{array}{cccccccc}
    \tilde{\textbf{h}}^R_{1}\\
    \tilde{\textbf{h}}^I_{1}\\
    \vdots \\
    \tilde{\textbf{h}}^R_{N_r}\\
    \tilde{\textbf{h}}^I_{N_r}
\end{array}
\right],~  \mathscr{A}_i=\left[
\begin{array}{cccccccc}
    \mathcal {A}_i   &   \textbf{0} & \cdots & \textbf{0}\\
    \textbf{0}   &   \mathcal {A}_i & \cdots & \textbf{0}\\
    \vdots & \vdots & \ddots & \vdots \\
    \textbf{0} & \textbf{0} & \cdots & \mathcal {A}_i
\end{array}
\right]_{N_r\times N_r},\\&~~~~~~~~~~ \mathscr{B}_i=\left[
\begin{array}{cccccccc}
    \mathcal {B}_i   &   \textbf{0} & \cdots & \textbf{0}\\
    \textbf{0}   &   \mathcal {B}_i & \cdots & \textbf{0}\\
    \vdots & \vdots & \ddots & \vdots \\
    \textbf{0} & \textbf{0} & \cdots & \mathcal {B}_i
\end{array}
\right]_{N_r\times N_r}.
\end{split}
\end{equation*}
Due to (\ref{bostc_condition2}), we have
$\mathscr{A}^T_i\mathscr{A}_i=\textbf{I},~\mathscr{B}^T_i\mathscr{B}_i=\textbf{I}~
(i=1,\cdots,k)$, and
$|\textbf{h}_1|=|\textbf{h}_2|=\cdots=|\textbf{h}_{2k}|=
|\bar{\textbf{h}}|$;

Due to (\ref{bostc_condition3}), an STC with dispersion matrices $\textbf{A}_1,\cdots,\textbf{A}_k$ are orthogonal and hence its equivalent channel matrix $\textbf{H}_1$ satisfies $\textbf{H}_1^T\textbf{H}_1=|\bar{\textbf{h}}|^2\textbf{I}$ (a detailed proof can be found in \cite{2gp_Yuen}). Similarly, due to (\ref{bostc_condition4}), we have $\textbf{H}_2^T\textbf{H}_2=|\bar{\textbf{h}}|^2\textbf{I}$. Under QR decomposition, $\textbf{H}=\textbf{QR}$ with
$\textbf{Q}\triangleq[\textbf{Q}_1~\textbf{Q}_2]$,
$\textbf{Q}_1\triangleq[\textbf{q}_1~\cdots~
\textbf{q}_k]=\frac{1}{|\bar{\textbf{h}}|}\textbf{H}_1$ and
$\textbf{Q}_2\triangleq[\textbf{q}_{k+1}~\cdots~\textbf{q}_{2k}]$;
$\textbf{R}\triangleq\left[
\begin{array}{cc}   \textbf {R}_1   &   \textbf{E}    \\
                    \textbf{0}      &   \textbf {R}_2
\end{array}
\right]$ is full-rank due to (\ref{bostc_condition1}); $\textbf
{R}_1=|\bar{\textbf{h}}|\textbf{I}_{k\times k}$ due to (\ref{bostc_condition3});
$\textbf{E}=\textbf{Q}^T_1\textbf{H}_2$.

In the following, we will prove that $\textbf {R}_2$ is diagonal and hence this STC has block-orthogonal structure $(2,k,1)$. We can see that
\begin{subequations}
\begin{align*}
\textbf{H}_2&=\textbf{Q}_1\textbf{E}+\textbf{Q}_2\textbf{R}_2\\
\textbf{H}_2-\textbf{Q}_1\textbf{E}&=\textbf{Q}_2\textbf{R}_2\\
(\textbf{H}_2-\textbf{Q}_1\textbf{E})^T(\textbf{H}_2-\textbf{Q}_1\textbf{E})
&=\textbf{R}^T_2\textbf{Q}^T_2\textbf{Q}_2\textbf{R}_2~\\&(\emph{\emph{where }}\textbf{Q}^T_2\textbf{Q}_2=\textbf{I}) \\
|\bar{\textbf{h}}|^2\textbf{I}+\textbf{E}^T\textbf{E}-\textbf{H}^T_2\textbf{Q}_1\textbf{E}-\textbf{E}^T\textbf{Q}^T_1\textbf{H}_2&
=\textbf{R}^T_2\textbf{R}_2~\\&(\emph{\emph{where }}\textbf{Q}^T_1\textbf{H}_2=\textbf{E})\\
|\bar{\textbf{h}}|^2\textbf{I}-\textbf{E}^T\textbf{E}&=\textbf{R}^T_2\textbf{R}_2
\end{align*}
\end{subequations}

In other words, $\textbf{E}^T\textbf{E}$ is diagonal $\Leftrightarrow~\textbf{R}^T_2\textbf{R}_2$ is diagonal.
Since $\textbf{R}_2$ is upper triangular, $\textbf{R}^T_2\textbf{R}_2$ is diagonal $\Leftrightarrow~\textbf{R}_2$ is diagonal. Hence, in the following, we will prove that $\textbf{E}^T\textbf{E}$ is diagonal under the condition (\ref{bostc_condition}).

Since $\textbf{E}=\textbf{Q}^T_1\textbf{H}_2$, we have
\begin{equation*}
\begin{split}
\textbf{E}&=\frac{1}{|\bar{\textbf{h}}|}\textbf{H}^T_1\textbf{H}_2\\
\textbf{E}^T\textbf{E}&=\frac{1}{|\bar{\textbf{h}}|^2}\textbf{H}^T_2\textbf{H}_1\textbf{H}^T_1\textbf{H}_2\\
&=\frac{1}{|\bar{\textbf{h}}|^2}\left[\textbf{h}^T_{k+i}\textbf{H}_1\textbf{H}^T_1\textbf{h}_{k+j}\right]\\
&=\frac{1}{|\bar{\textbf{h}}|^2}\left[\bar{\textbf{h}}^T\mathscr{B}^T_{i}\textbf{H}_1
\textbf{H}^T_1\mathscr{B}_{j}\bar{\textbf{h}}\right]
\end{split}
\end{equation*}

To ensure that
$\textbf{E}^T\textbf{E}=\frac{1}{|\bar{\textbf{h}}|^2}\left[\bar{\textbf{h}}^T\mathscr{B}^T_{i}\textbf{H}_1
\textbf{H}^T_1\mathscr{B}_{j}\bar{\textbf{h}}\right]$ is diagonal,
we need
\begin{equation} \label{eq_EEdiagcondition}
\begin{split}
&\bar{\textbf{h}}^T\mathscr{B}^T_{i}\textbf{H}_1\textbf{H}^T_1\mathscr{B}_{j}\bar{\textbf{h}}=0,~i,j=1,\cdots,k,~i\neq j.
\end{split}
\end{equation}

Let
$\mathscr{A}_{i}\triangleq\left[\mathbbm{a}_{i1}~\mathbbm{a}_{i2}~\cdots~\mathbbm{a}_{i2TN_r}\right]^T
\triangleq[a_{iuv}]_{2TN_r\times 2N_tN_r}$ and
$\mathscr{B}_{i}\triangleq\left[\mathbbm{b}_{i1}~\mathbbm{b}_{i2}~\cdots~\mathbbm{b}_{i2N_tN_r}\right]
\triangleq[b_{iuv}]_{2TN_r\times 2N_tN_r}$, we have
\begin{equation*}
\begin{split}
\mathscr{B}^T_{i}\textbf{H}_1\textbf{H}^T_1\mathscr{B}_{j}&\triangleq[w_{pq}]_{2N_tN_r\times
2N_tN_r}\\
&=\mathscr{B}^T_{i}\left[\mathscr{A}_{1}\bar{\textbf{h}}~\cdots~
\mathscr{A}_{k}\bar{\textbf{h}}\right]
\\&~~~\left[\mathscr{A}_{1}\bar{\textbf{h}}~\cdots~
\mathscr{A}_{k}\bar{\textbf{h}}\right]^T\mathscr{B}_{j}\\
\end{split}
\end{equation*}
\noindent with
\begin{equation*}
\begin{split}
w_{pq}&=\mathbbm{b}^T_{ip}\left[\mathscr{A}_{1}\bar{\textbf{h}}~\cdots~
\mathscr{A}_{k}\bar{\textbf{h}}\right]
\left[\mathscr{A}_{1}\bar{\textbf{h}}~\cdots~
\mathscr{A}_{k}\bar{\textbf{h}}\right]^T\mathbbm{b}_{jq}\\
&=\left[\sum^{2TN_r}_{u=1}b_{iup}\mathbbm{a}_{1u}\bar{\textbf{h}}~\cdots~
\sum^{2TN_r}_{u=1}b_{iup}\mathbbm{a}_{ku}\bar{\textbf{h}}\right]\\&~~~~\cdot
\left[\sum^{2TN_r}_{v=1}b_{ivq}\mathbbm{a}_{1v}\bar{\textbf{h}}~\cdots~
\sum^{2TN_r}_{v=1}b_{jvq}\mathbbm{a}_{kv}\bar{\textbf{h}}\right]^T\\
&=\sum^{k}_{\kappa=1} \sum^{2TN_r}_{u=1}b_{iup}\mathbbm{a}_{\kappa
u}\bar{\textbf{h}}\cdot
\sum^{2TN_r}_{v=1}b_{jvq}\mathbbm{a}_{\kappa v}\bar{\textbf{h}}\\
&=\bar{\textbf{h}}^T\sum^{k}_{\kappa=1}
\left(\sum^{2TN_r}_{u=1}b_{iup}\mathbbm{a}^T_{\kappa u}\cdot
\sum^{2TN_r}_{v=1}b_{jvq}\mathbbm{a}_{\kappa v}\right)\bar{\textbf{h}}.\\
\end{split}
\end{equation*}

For a clear presentation, we define
\begin{equation*}
\begin{split}
& \sum^{k}_{\kappa=1}\left(
\sum^{2TN_r}_{u=1}b_{iup}\mathbbm{a}^T_{\kappa u}\cdot
\sum^{2TN_r}_{v=1}b_{jvq}\mathbbm{a}_{\kappa v}\right)\\
\triangleq &~~\textbf{D}_{pq}\triangleq [d_{pqst}]_{2N_tN_r\times 2N_tN_r}
\end{split}
\end{equation*}
where
\begin{equation*}
\begin{split}
d_{pqst}=\sum^{k}_{\kappa=1}\left(
\sum^{2TN_r}_{u=1}b_{iup}a_{\kappa us}\cdot\right.\left.
\sum^{2TN_r}_{v=1}b_{jvq}a_{\kappa vt}\right).
\end{split}
\end{equation*}

With $i,j=1,\cdots,k,i\neq j$ and
$\bar{\textbf{h}}=[\bar{h}_1~\cdots~\bar{h}_{2N_tN_r}]^T$, for condition (\ref{eq_EEdiagcondition}) to be valid, we first simplified the term $\bar{\textbf{h}}^T\mathscr{B}^T_{i}\textbf{H}_1\textbf{H}^T_1\mathscr{B}_{j}\bar{\textbf{h}}$ as follows:
\begin{equation} \label{eq_EEdiagcondition_term}
\begin{split}
&\bar{\textbf{h}}^T\mathscr{B}^T_{i}\textbf{H}_1\textbf{H}^T_1\mathscr{B}_{j}\bar{\textbf{h}}\\
=&\bar{\textbf{h}}^T [w_{pq}]_{2N_tN_r\times 2N_tN_r}
\bar{\textbf{h}}\\
=&\sum^{2N_tN_r}_{p=1}\bar{h}_p\sum^{2N_tN_r}_{q=1}\bar{h}_qw_{pq}\\
=&\sum^{2N_tN_r}_{p=1}\bar{h}_p\sum^{2N_tN_r}_{q=1}\bar{h}_q\cdot
\bar{\textbf{h}}^T\\&~~~~~\sum^{k}_{\kappa=1}\left(
\sum^{2TN_r}_{u=1}b_{iup}\mathbbm{a}^T_{\kappa u}\cdot
\sum^{2TN_r}_{v=1}b_{jvq}\mathbbm{a}_{\kappa v}\right)\bar{\textbf{h}}\\
=&\sum^{2N_tN_r}_{p=1}\bar{h}_p \sum^{2N_tN_r}_{q=1}\bar{h}_q
\sum^{2N_tN_r}_{s=1}\bar{h}_s \sum^{2N_tN_r}_{t=1}\bar{h}_t \cdot
d_{pqst}.
\end{split}
\end{equation}

Since $\bar{h}_p$, $\bar{h}_q$, $\bar{h}_s$ and $\bar{h}_t$ are random channel coefficients, for $\bar{\textbf{h}}^T\mathscr{B}^T_{i}\textbf{H}_1\textbf{H}^T_1\mathscr{B}_{j}\bar{\textbf{h}}$, i.e., (\ref{eq_EEdiagcondition_term}), being 0, all the coefficients of the polynomial $\sum^{2N_tN_r}_{q=1}\bar{h}_q
\sum^{2N_tN_r}_{p=1}\bar{h}_p \sum^{2N_tN_r}_{s=1}\bar{h}_s \sum^{2N_tN_r}_{t=1}\bar{h}_t
\cdot d_{pqst}$ should be 0, i.e.,
\begin{equation} \label{bostc_condition5_o}
\begin{split}
\sum_{(p,q,s,t)\in \mathbb{S}_0} d_{pqst}=0
\end{split}
\end{equation}
\noindent where each element (tuple) of set $\mathbb{S}_0$ includes 4 uniquely-permuted
scalars\footnote{For example, $\sum_{(1,2,1,1)\in \mathbb{S}_0} d_{pqst}=
d_{1112}+d_{1121}+d_{1211}+d_{2111}$ and $\sum_{(1,2,3,1)\in \mathbb{S}_0} d_{pqst}=
d_{1123}+d_{1132}+d_{1213}+d_{1312}+d_{1231}+d_{1321}+d_{2113}+d_{2131}+d_{2311}+d_{3112}+d_{3121}+d_{3211}$.} drawn from
$\{1,\cdots,2N_tN_r\}$ and corresponds to a term $\bar{h}_p\bar{h}_q\bar{h}_s\bar{h}_t$ with coefficient
$\sum_{(p,q,s,t)\in \mathbb{S}_0} d_{pqst}$.

Since $\mathscr{A}_{\kappa}(\mathscr{B}_{i})$ is block-diagonal
with the same main diagonal sub-matrix $\mathcal {A}_{\kappa}(\mathcal
{B}_{i},~\kappa,i=1,\cdots,k)$, there must be at least one 0 value between $b_{iup}$ and
$a_{\kappa us}$, i.e., $b_{iup}a_{\kappa us}=0$, when $p$ and $s$ correspond to
two diagonal sub-matrices, i.e., $\lfloor \frac{p}{N_r}\rfloor\neq
\lfloor \frac{s}{N_r}\rfloor$ with the floor function $\lfloor \cdot
\rfloor$. Hence, $p\emph{\emph{ and }}s$ can be considered to be corresponding to the same sub-matrix $\mathcal {A}_{\kappa}(\mathcal {B}_{i})$. Hence (\ref{bostc_condition5_o}) is equivalent to (\ref{bostc_condition5_s}):
\begin{equation} \label{bostc_condition5_s}
\begin{split}
\sum_{(p,q,s,t)\in \mathbb{S}} d_{pqst}=0
\end{split}
\end{equation}
\noindent where each element (tuple) of set $\mathbb{S}$ includes 4
uniquely-permuted scalars drawn from $\{1,\cdots,2N_t\}$.

Hence, with (\ref{bostc_condition}), $\textbf{E}^T\textbf{E}=\frac{1}{|\bar{\textbf{h}}|^2}\left[\bar{\textbf{h}}^T\mathscr{B}^T_{i}\textbf{H}_1
\textbf{H}^T_1\mathscr{B}_{j}\bar{\textbf{h}}\right]$ is diagonal, i.e., $\textbf{R}_2$ is diagonal. Since $\textbf{R}_1$ and $\textbf{R}_2$ are diagonal, Theorem (\ref{th_bostbc_a}) is proved.

\section{} \label{proof_th_bostbc_b}

Since $\{\textbf{B}_{1},\cdots, \textbf{B}_{k}\}$ satisfy the QOC, $\textbf{H}^T_2\textbf{H}_2$ is diagonal.

Under QR decomposition,
\begin{equation} \label{eq_qronH_borelation}
\begin{split}
\textbf{H}=\left[\textbf{H}_1~\textbf{H}_2\right]=\textbf{QR}=\left[\textbf{Q}_1~\textbf{Q}_2\right]
\left[
\begin{array}{ccccc}
    \textbf{R}_1   &   \textbf{E}_{12}\\
    \textbf{0}     &   \textbf{R}_2
\end{array}
\right]
\end{split}
\end{equation}
\noindent where $\textbf{E}_{12}$ is the projection coefficient matrix of vectors $\textbf{h}_{\mathbbm{k}+1},\cdots,\textbf{h}_{\mathbbm{k}+k}$ onto vector space $\{\textbf{h}_1,\cdots,\textbf{h}_{\mathbbm{k}}\}$. Following the QR decomposition algorithm, we see that $\textbf{E}_{12}=\textbf{Q}^T_1\textbf{H}_2$.

In (\ref{eq_qronH_borelation}), we have
\begin{equation*}
\begin{split}
\textbf{H}_2&=\textbf{Q}_1\textbf{E}_{12}+\textbf{Q}_2\textbf{R}_2 \\
\textbf{H}_2-\textbf{Q}_1\textbf{E}_{12}&=\textbf{Q}_2\textbf{R}_2 \\
(\textbf{H}_2-\textbf{Q}_1\textbf{E}_{12})^T(\textbf{H}_2-\textbf{Q}_1\textbf{E}_{12})
&=(\textbf{Q}_2\textbf{R}_2)^T(\textbf{Q}_2\textbf{R}_2) \\
\textbf{H}^T_2\textbf{H}_2-\textbf{E}^T_{12}\textbf{E}_{12}&=\textbf{R}^T_2\textbf{R}_2~
(\textbf{Q}^T_2\textbf{Q}_2=\textbf{I}) \\
\end{split}
\end{equation*}

Hence, with diagonal $\textbf{H}^T_2\textbf{H}_2$, we have: $\textbf{E}^T\textbf{E}$ is diagonal $\Leftrightarrow$
$~\textbf{R}^T_2\textbf{R}_2$ is diagonal $\Leftrightarrow$ $\textbf{R}_2$ is diagonal, where $\textbf{R}_2$ has been known to be upper triangular.

Hence Theorem \ref{th_bostbc_b} is proved.

\begin{biography}
[{\includegraphics[width=1in,height=1.25in,clip,keepaspectratio]{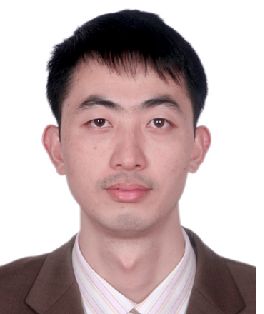}}]{Tian Peng Ren}
received the Bachelor of Engineering from Air Force Engineering University, Xi\textquoteright an, China, in 2004, and the M.S. and Ph.D. degrees of Electrical Engineering from National University of Defense Technology (NUDT), Changsha, China, in 2007 and 2010 respectively. He is now with 63790 troops, Xichang 615000, China. His current research interests lie in the area of wireless communications, including MIMO systems and space-time coding.
\end{biography}

\begin{biography}
[{\includegraphics[width=1in,height=1.25in,clip,keepaspectratio]{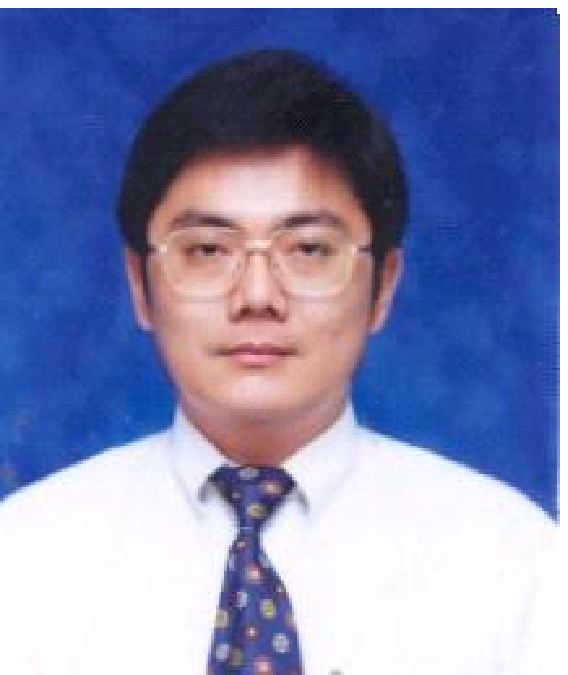}}]{Yong Liang Guan}
obtained his Ph.D. degree from the Imperial College of London, UK, in 1997 and Bachelor of Engineering with first class honors from the National University of Singapore in 1991. He is now an Associate Professor with the School of Electrical and Electronic Engineering, and the Director of the Positioning and Wireless Technology Center (http://www3.ntu.edu.sg/Centre/pwtc/profile.html), at the Nanyang Technological University of Singapore. He has also been appointed an Adjunct Professor of the University of Electronic Science and Technology of China, Chengdu, China, and a Faculty Associate of the Institute of Infocomm Research, Agency of Science, Technology and Research, Singapore. His research interests broadly include modulation, coding and signal processing for communication systems and information security systems. He is an associate editor of the IEEE Signal Processing Letters. His homepage is at http://www3.ntu.edu.sg/home/eylguan/index.htm.
\end{biography}

\begin{biography}
[{\includegraphics[width=1in,height=1.25in,clip,keepaspectratio]{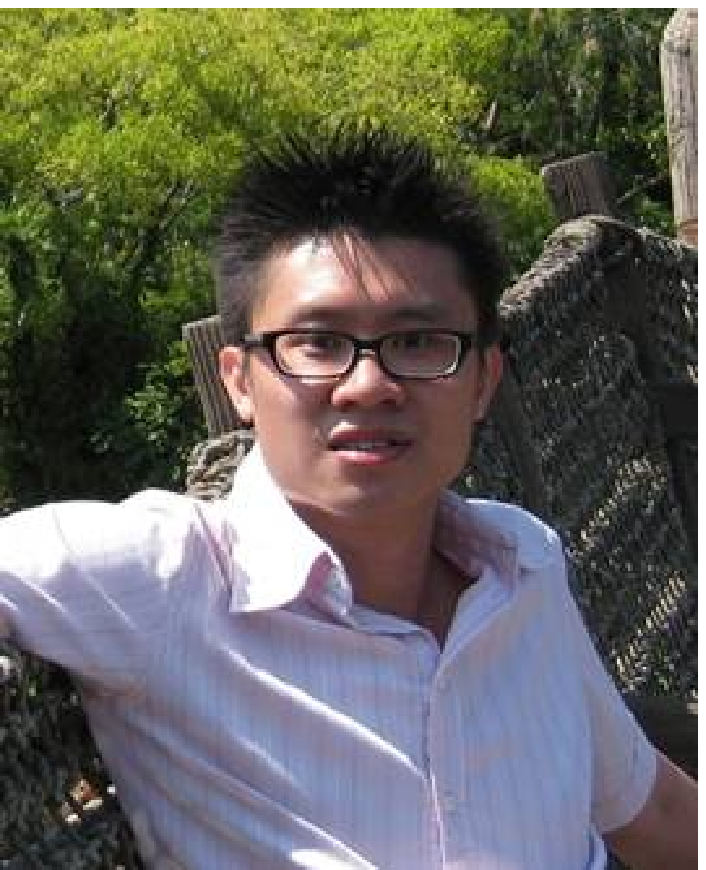}}]{Chau Yuen}
received the Bachelor of Engineering and Ph.D. degree from Nanyang Technological University (NTU), Singapore in 2000 and 2004, respectively. He is the recipient of Lee Kuan Yew Gold Medal, Institution of Electrical Engineers (IEE) Book Prize, Institute of Engineering of Singapore (IES) Gold Medal, Merck Sharp \& Dohme (MSD) Gold Medal and twice the recipient of Hewlett Packard (HP) Prize. He was a Post Doctor Fellow in Lucent Technologies Bell Labs, Murray Hill during 2005. Dr. Yuen was a Visiting Assistant Professor of Hong Kong Polytechnic University in 2008. During 2006-2010, he worked at Institute for Infocomm Research (I2R, Singapore) as a Senior Research Engineer, where he involved in an industrial project on developing an 802.11n Wireless LAN system, and participated actively in 3Gpp Long Term Evolution (LTE) and LTE-Advanced (LTE-A) standardization. He serves as an Associate Editor for IEEE Transactions on Vechicular Technology. He has published over 70 research papers at international journals or conferences. His present research interests include green communications, cooperative transmissions, network coding, wireless positioning, and wireless network. He joined Singapore University of Technology and Design as an Assistant Professor from June 2010.
\end{biography}

\begin{biography}
[{\includegraphics[width=1in,height=1.25in,clip,keepaspectratio]{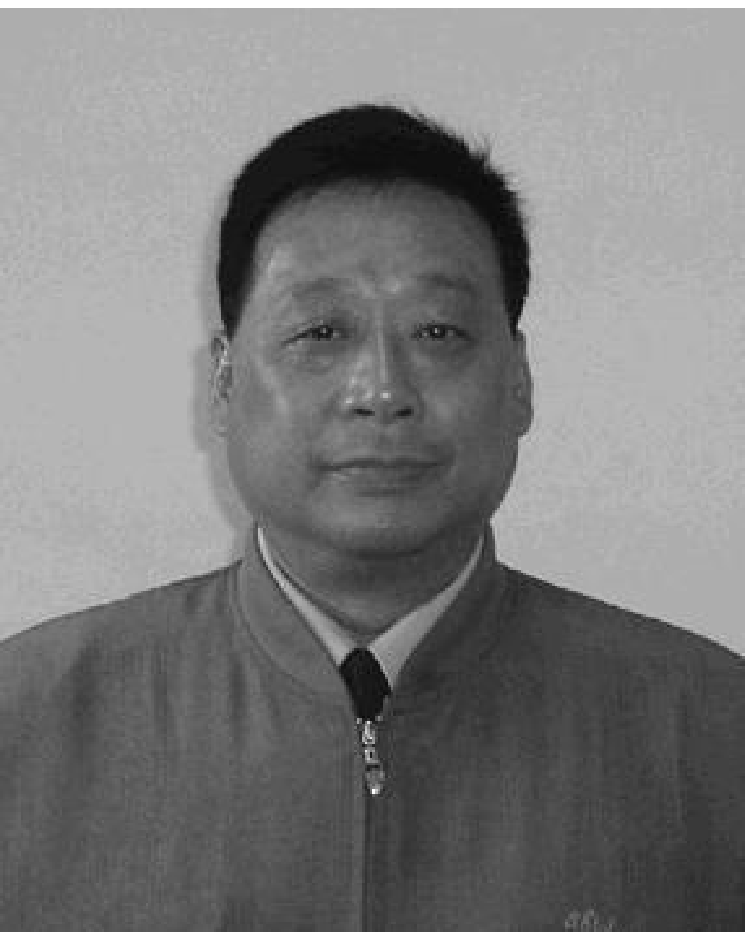}}]{Er Yang Zhang}
received the Bachelor of Engineering from the Military Academy of Engineering, Harbin, China, in 1964.

Currently, he is a Full Professor at National University of Defense Technology (NUDT), Changsha, China. He is a Fellow of the China Institute of Communications, and a Senior Member of the Chinese Institute of Electronics. From 1984 to 1987, he was with the University of Notre Dame, Chicago, as a Visiting Scholar. He was a committee number of National 863 Project in the aerospace field during 1993-2001. His research interests include wireless communications, satellite communications and network technologies.
\end{biography}


\begin{thebibliography}{100}

\bibitem{Alamouti}S. M. Alamouti, \textquoteleft \textquoteleft A simple transmitter diversity scheme for wireless communication,\textquoteright\textquoteright~\emph{IEEE J. Sel. Areas Commun.}, vol.16, pp. 1451-1458, Oct. 1998.

\bibitem{Tarokh_ostbc}V. Tarokh, H. Jafarkhani, and A. R. Calderbank, \textquoteleft \textquoteleft Space-time block codes from orthogonal designs,\textquoteright\textquoteright~\emph{IEEE Trans. Inf. Theory}, vol. 45, no. 5, pp. 1456-1466, Jul. 1999.

\bibitem{Ganesan_ostbc}G. Ganesan and P. Stoica, \textquoteleft \textquoteleft Space-time block codes: A maximum SNR approach,\textquoteright\textquoteright~\emph{IEEE Trans. Inf. Theory}, vol. 47, no. 4, pp. 1650-1656, May 2001.

\bibitem{Lu_ostbc}K. Lu, S. Fu, and X.-G. Xia, \textquoteleft \textquoteleft Closed-form designs of complex orthogonal space-time block codes of rates $(k+1)/(2k)$ for $2k-1$ or $2k$ transmit antennas ,\textquoteright\textquoteright~\emph{IEEE Trans. Inf. Theory}, vol. 51, no. 12, pp. 4340-4347, Dec. 2005.

\bibitem{Wang_upperbound}H. Wang and X.-G. Xia, \textquoteleft \textquoteleft Upper bounds of rates of complex orthogonal space-time block codes,\textquoteright\textquoteright~\emph{IEEE Trans. Inf. Theory}, vol. 49, no. 10, pp. 2788-2796, Oct. 2003.

\bibitem{qostbc_Jafarkhani}H. Jafarkhani, \textquoteleft \textquoteleft A quasi-orthogonal space-time block code, \textquoteright\textquoteright~\emph{IEEE Trans. Commun.}, vol. 49, no. 1, pp. 1-4, Jan. 2001.

\bibitem{qostbc_Tirkkonen}O. Tirkkonen, A. Boariu and A. Hottinen, \textquoteleft \textquoteleft Minimal non-orthogonality rate 1 space-time block code for 3+ Tx antennas,\textquoteright\textquoteright~in~\emph{Proc. IEEE ISSSTA}, Parsippany, NJ, Sept. 6-8, 2000.

\bibitem{qostbc_Papadias}C. B. Papadias and G. J. Foschini, \textquoteleft \textquoteleft A space-time coding approach for systems employing four transmit antennas,\textquoteright\textquoteright~in~\emph{Proc. IEEE ICASSP}, Salt Lake City, UT, 2001.

\bibitem{qostbc_Mecklenbrauker}C. F. Mecklenbrauker and M. Rupp, \textquoteleft \textquoteleft Generalized Alamouti codes for trading quality of service against data rate in MIMO UMTS, \textquoteright\textquoteright~\emph{EURASIP J. Appl. Signal Processing}, no. 5, pp. 662¨C675, May 2004.

\bibitem{qostbc_Yuen}C. Yuen, Y. Guan, and T. T. Tjhung, \textquoteleft \textquoteleft Quasi-orthogonal STBC with minimum decoding complexity, \textquoteright\textquoteright~\emph{IEEE Trans. Wireless Commun.}, vol. 4, pp. 2089¨C2094, Sep. 2005.

\bibitem{qostbc_Rajan}Z. A. Khan and B. S. Rajan, \textquoteleft \textquoteleft Single-symbol maximum-likelihood decodable linear STBCs,\textquoteright\textquoteright~\emph{IEEE Trans. Inf. Theory}, vol. 52, pp. 2062¨C2091, May 2006.

\bibitem{qostbc_Dao}D. N. Dao, C. Yuen, C. Tellambura, Y. L. Guan and T. T. Tjhung, \textquoteleft \textquoteleft Four-group decodable space-time block codes,\textquoteright\textquoteright~\emph{IEEE Trans. Signal Process.}, vol. 56, no. 1, pp. 424-430, Jan. 2008.

\bibitem{2gp_Yuen}C. Yuen, Y. L. Guan, and T. T. Tjhung, \textquoteleft \textquoteleft On the search for high-rate quasi-orthogonal space-time block code,\textquoteright\textquoteright~\emph{Int. J. Wireless Information Network (IJWIN)}, vol. 13, pp. 329-340, Oct. 2006.

\bibitem{2gp_Rajan}K. Pavan Srinath and B. Sundar Rajan, \textquoteleft \textquoteleft High-rate, 2-group ML-decodable STBCs for $2^m$ transmit antennas,\textquoteright\textquoteright~in~\emph{Proc. IEEE ISIT\textquoteright09}, Seoul, Korea, June 28-July 3 2009.

\bibitem{2gp_Ren}T. P. Ren, Y. L. Guan, C. Yuen, E. Gunawan and E. Y. Zhang, \textquoteleft \textquoteleft Group-Decodable Space-Time Block Codes with Code Rate $>$ 1,\textquoteright\textquoteright~accepted by \emph{IEEE Trans. Commun.}. See also \textquoteleft \textquoteleft Unbalanced and balanced 2-group decodable spatial multiplexing code,\textquoteright\textquoteright~in~\emph{Proc. IEEE VTC\textquoteright09-Fall}, Anchorage, Alaska, 20-23 Sept. 2009.

\bibitem{Foschini}G. J. Foschini, \textquoteleft \textquoteleft Layered space time architecture for wireless communication in a fading environment when using multi-element antennas,\textquoteright\textquoteright~\emph{Bell labs Tech. Journal}, vol. 1, pp. 41-59, 1996.

\bibitem{Texas_dsttd}Texas Instruments, \textquoteleft \textquoteleft Improved double-STTD scheme using asymmetric modulation and antenna shuffling,\textquoteright\textquoteright~3GPP TSGR1\#20(01)-0459, Busan, Korea, May 21-25, 2001.

\bibitem{Hottinen}A. Hottinen, R. Wichman, and O. Tirkkonen, \emph{Multiantenna Transceiver Techniques for 3G and Beyond},~John Wiley and Sons, Feb. 2003.

\bibitem{Belfiore}J.-C. Belfiore, G. Rekaya, and E. Viterbo, \textquoteleft \textquoteleft The Golden code: A 2x2 full-rate space-time code with non-vanishing determinants,\textquoteright\textquoteright~\emph{IEEE Trans. Inf. Theory}, vol. 51, pp. 1432-1436, Apr. 2005.

\bibitem{Oggier}F. Oggier, G. Rekaya, J. C. Belfiore, and E. Viterbo, \textquoteleft \textquoteleft Perfect space-time block codes,\textquoteright\textquoteright~\emph{IEEE Trans. Inf. Theory}, vol. 52, no. 9, pp. 3885-3902, 2006.

\bibitem{Biglieri}E. Biglieri, Y. Hong, E. Viterbo, \textquoteleft \textquoteleft On fast-decodable space-time block codes,\textquoteright\textquoteright~\emph{IEEE Trans. Inf. Theory}, vol. 55, no. 2, pp. 524-530, Feb. 2009.

\bibitem{Ren_fgd}T. P. Ren, Y. L. Guan, C. Yuen and R. J. Shen, \textquoteleft \textquoteleft Fast-group-decodable space-time block code,\textquoteright\textquoteright~in~\emph{Proc. IEEE ITW\textquoteright10}, Cairo, Egypt, 6-8 Jan. 2010.

\bibitem{Hassibi}B. Hassibi and B. M. Hochwald, \textquoteleft \textquoteleft High-rate codes that are linear in space and time,\textquoteright\textquoteright~\emph{IEEE Trans. Inf. Theory}, vol. 48, no. 7, pp. 1804-1824, Jul. 2002.

\bibitem{Yuen_book}C. Yuen, Y. L. Guan, and T. T. Tjhung, \emph{Quasi-orthogonal space-time block code},~Imperial College Press, 2007.

\bibitem{ren_bostc}T. P. Ren, Y. L. Guan, C. Yuen and E. Y. Zhang, \textquoteleft \textquoteleft Block-orthogonal space-time codes with decoding complexity reduction,\textquoteright\textquoteright~in~\emph{Proc. IEEE SPAWC}, Marrakech, Morocco, June 20-23, 2010.

\bibitem{Damen}M. O. Damen, H. El Gamal, and G. Caire, \textquoteleft \textquoteleft On maximum-likelihood detection and the search for the closest lattice point, \textquoteright\textquoteright~\emph{IEEE Trans. Inf. Theory}, vol. 49, no. 10, pp. 2389-2402, Oct. 2003.

\bibitem{Kim}K. J. Kim and R. A.Iltis, \textquoteleft \textquoteleft Joint detection and channel estimation algorithms for QS-CDMA signals over time-varying channels,\textquoteright\textquoteright~\emph{IEEE Trans. Commun.}, vol.50, no.5, pp. 845-855, May. 2002.

\bibitem{Chin}W. H. Chin, \textquoteleft \textquoteleft QRD based tree search data detection for MIMO communication systems,\textquoteright\textquoteright~in~\emph{Proc. IEEE VTC\textquoteright05-Spring}, Stockholm, Sweden, 2005, pp. 1624-1627.

\bibitem{Paredes}J. Paredes, A. B. Gershman, and M. G. Alkhanari, \textquoteleft \textquoteleft A new full-rate full-diversity space-time block code with nonvanishing determinants and simplified maximum-likelihood decoding,\textquoteright\textquoteright~\emph{IEEE Trans. Signal Process.}, vol. 56, pp. 2461-2469, June 2008.

\bibitem{Sezginer}S. Sezginer, H. Sari and E. Biglieri, \textquoteleft \textquoteleft On high-rate full-diversity 2$\times$2 space-time codes with low-complexity optimum detection,\textquoteright\textquoteright~\emph{IEEE Trans. Commum.}, vol. 57, no. 5, pp. 1532-1541, May 2009.

\bibitem{Rabiei}P. Rabiei, N. Al-Dhahir and R. Calderbank, \textquoteleft \textquoteleft New rate-2 STBC design for 2 TX with reduced-complexity maximum likelihood decoding,\textquoteright\textquoteright~\emph{IEEE Trans. Wireless Commun.}, vol. 8, no. 4, pp. 1803-1813, Apr. 2009.

\bibitem{Thompson}A. R. Thompson, J. M. Moran and G. W. Jr Swenson, \emph{Interferometry and Synthesis in Radio Astronomy},~New York: Wiley, pp. 204, 1986.

\bibitem{codeexample_matlab} http://www.pwtc.eee.ntu.edu.sg/Research/Documents/Code\_verify\_J \_STSP\_SDWT\_00024\_2011.m

\bibitem{Ren_bostr}T. P. Ren, Y. L. Guan, C. Yuen and E. Y. Zhang, \textquoteleft \textquoteleft Space-Time Codes with Block-Orthogonal Structure and Their Simplified ML and Near-ML Decoding,\textquoteright\textquoteright~in~\emph{Proc. IEEE VTC\textquoteright10-Fall}, Ottawa, Canada, 6-9 Sept. 2010.

\end{thebibliography}
\end{document}